\documentclass[
 reprint,
superscriptaddress,
 amsmath,amssymb,
prl,
floatfix,
]{revtex4-2}

\usepackage{graphicx}
\usepackage{dcolumn}
\usepackage{bm}
\usepackage{color}
\usepackage[normalem]{ulem}

\usepackage[colorlinks,urlcolor=blue,citecolor=blue,linkcolor=blue]{hyperref}

\begin{document}

\preprint{topo}





\title{Negative-mass exciton polaritons induced by dissipative light-matter coupling in an atomically thin semiconductor}

\author{M.~Wurdack}
\email{matthias.wurdack@anu.edu.au}
\affiliation{ARC Centre of Excellence in Future Low-Energy Electronics Technologies and Department of Quantum Science and Technology, Research School of Physics, The Australian National University, Canberra, ACT 2601, Australia}

\author{T.~Yun}%
\affiliation{ARC Centre of Excellence in Future Low-Energy Electronics Technologies and Department of Quantum Science and Technology, Research School of Physics, The Australian National University, Canberra, ACT 2601, Australia}
\affiliation{Department of Materials Science and Engineering, Monash University, Clayton, Victoria, 3800, Australia}
\affiliation{{\color{black}Songshan Lake Materials Laboratory, Dongguan 523808, Guangdong, China}}
\affiliation{{\color{black}Institute of Physics, Chinese Academy of Science, Beijing 100190, China}}

\author{M.~Katzer}
\affiliation{Nichtlineare Optik und Quantenelektronik, Institut f\"ur Theoretische Physik, Technische Universit\"at Berlin,  10623 Berlin, Germany}

\author{A.~G.~Truscott}%
\affiliation{Department of Quantum Science and Technology, Research School of Physics, The Australian National University, Canberra, ACT 2601, Australia}

\author{A.~Knorr}
\affiliation{Nichtlineare Optik und Quantenelektronik, Institut f\"ur Theoretische Physik, Technische Universit\"at Berlin,  10623 Berlin, Germany}

\author{M.~Selig}
\affiliation{Nichtlineare Optik und Quantenelektronik, Institut f\"ur Theoretische Physik, Technische Universit\"at Berlin,  10623 Berlin, Germany}
  
\author{E.~A.~Ostrovskaya}
\email{elena.ostrovskaya@anu.edu.au}
\affiliation{ARC Centre of Excellence in Future Low-Energy Electronics Technologies and Department of Quantum Science and Technology, Research School of Physics, The Australian National University, Canberra, ACT 2601, Australia}

\author{E.~Estrecho}%
\email{eliezer.estrecho@anu.edu.au}
\affiliation{ARC Centre of Excellence in Future Low-Energy Electronics Technologies and Department of Quantum Science and Technology, Research School of Physics, The Australian National University, Canberra, ACT 2601, Australia}

\maketitle 


\textbf{Dispersion engineering is a powerful and versatile tool that can vary the speed of light signals and induce negative-mass effects in the dynamics of {\color{black}particles and quasiparticles}. Here, we show that dissipative coupling {\color{black}between} bound electron-hole pairs (excitons) and photons in an optical microcavity can lead to the formation of exciton polaritons with an inverted dispersion of the lower polariton branch and hence a negative mass. We perform direct measurements of the anomalous dispersion in atomically thin (monolayer) WS$_2$ {\color{black} crystals} embedded in planar {\color{black} microcavities} and demonstrate that the propagation direction of the negative-mass polaritons is opposite to their momentum. Our study introduces a new concept of non-Hermitian dispersion engineering for exciton polaritons and opens a pathway for realising new phases of quantum matter in a solid state.}

Losses are ubiquitous in nature and are usually perceived as detrimental to the performance of electronic and photonic devices. However, recent understanding of the physics of non-Hermitian systems with loss and gain has led to the possibility of novel properties and functionalities by judicious control of losses. This concept is most powerfully demonstrated in non-Hermitian photonics \cite{El-Ganainy2019,Miri2019,Ozdemir2019}, where non-Hermitian spectral degeneracies (exceptional points) and associated symmetry-breaking transitions fundamentally change the laws of wave propagation and scattering.
Although the study of non-Hermitian physics in quantum electronic systems remains difficult, significant progress has been achieved in hybrid photonic systems, where photons are strongly coupled to electronic excitations in a solid state to form exciton polaritons, part-light part-matter hybrid quasiparticles \cite{Microcavities}. Nontrivial topology of the eigenstates \cite{Gao2015,Gao2018} and mode selectivity \cite{Li2021} in the vicinity of the exceptional points, band engineering \cite{Pickup2020} and nonlinear localisation \cite{Pernet2021} in non-Hermitian lattices, emergence of non-Hermitian topology \cite{SuEstrecho2020} and divergent quantum geometric metric near an exceptional point \cite{Liao2020} have been demonstrated in the strong light-matter coupling regime.



Here, we dramatically modify the exciton-polariton dispersion by exploiting a previously undetected non-Hermitian component of exciton-photon interaction called dissipative coupling~\cite{DissipativeCouplingOpto}. Our microscopic theory shows that this type of coupling can arise from the interplay of exciton-phonon scattering {\color{black} in monolayers of transition metal dichalcogenide crystals (TMDCs)} and photon losses. Also known as external coupling via the continuum~\cite{RotterReview}, dissipative coupling leads to level attraction or clustering~\cite{LevelAttractionMagnon,RotterReview} and resonance trapping~\cite{ResonanceTrappingExp} in other physical systems. This is in contrast to the well-known coherent (or internal) coupling which always leads to level repulsion. We show theoretically that the interplay between the coherent and dissipative light-matter coupling results in an inverted dispersion of exciton polaritons in a planar microcavity {\color{black} with an embedded monolayer TMDC}. We directly measure this anomalous dispersion {\color{black} in several planar microcavities with integrated monolayer WS$_2$ \cite{Yun2022} at room temperature, and demonstrate the negative-mass transport of exciton polaritons. {\color{black}The key role of the exciton-phonon scattering in the dissipative coupling mechanism is further confirmed by temperature-dependent measurements.}}


\section*{Results} 

\textbf{{\color{black}Theory.}}  To demonstrate the principle of non-Hermitian dispersion engineering, we start with an effective Hamiltonian given by $H=H_0 -iWW^\dagger$~\cite{ResonanceTrappingExp} describing the coherent (internal) and dissipative (external) coupling of cavity photons $|C\rangle$ and excitons $|X\rangle$:
\begin{equation}
\label{eq:Hamiltonian}
    H = \begin{pmatrix}
E_{c} & V\\ 
V & E_{x}
\end{pmatrix} -i
\begin{pmatrix}
\sqrt{\gamma_c} & \sqrt{g_x}\\ 
\sqrt{g_c} & \sqrt{\gamma_x}
\end{pmatrix}
\begin{pmatrix}
\sqrt{\gamma_c} & \sqrt{g_c}\\ 
\sqrt{g_x} & \sqrt{\gamma_x}
\end{pmatrix}.
\end{equation}
Here, the Hermitian term $H_0$ models the coherent coupling of excitons and photons with the bare energies $E_{c,x}$, respectively. The {\color{black}coherent} coupling strength $V$ is proportional to the exciton oscillator strength and the overlap of the exciton dipole with the confined electric field of the cavity photon~\cite{Microcavities}.
The matrix $W$ describes the external coupling to {\color{black} two dissipative channels in the system~\cite{ResonanceTrappingExp}: $\gamma_{c}$ is the coupling of cavity photons to the continuum of states outside the cavity due to the imperfect mirrors limiting their lifetimes, $\gamma_x$ is the coupling to radiative and non-radiative channels resulting in the homogeneous linewidth broadening of excitons~\cite{Selig2016,MoodyLinewidth2015}, and $g_{x,c}$ quantifies the effective dissipative coupling between excitons and cavity photons via the two decay channels.

This non-Hermitian, phenomenological Hamiltonian, Eq.~(1), is motivated by a full microscopic model (see Supplementary Information Section S1). In a self-consistent theory without free parameters, with all values from ab-initio calculations, we show that the off-diagonal dissipation terms in $W$ (cp. Eq.~(1))
arise due to the mixing of the two decay channels: (i) the dissipation of energy from the excitons via phonon scattering events in the TMDC, i.e. due to the phonon bath, leading to $g_x$ and (ii) the energy loss due to leakage of photons out of the cavity leading to $g_c$. Only for nonzero off-diagonal dissipation terms in Eq.~(1) one can observe level attraction and anomalous dispersion as described below.}

{\color{black}Remarkably, when the coupling to the phonon bath is `turned off' in the microscopic model, the 
effects arising from the dissipative coupling
disappear (see Supplementary Information). Since the exciton-phonon scattering is known to be significant in TMDCs ~\cite{MoodyLinewidth2015, Selig2016, Li2021Phonon, Li2022polariton}, we expect strong effects of dissipative coupling via this channel in this material system.
Furthermore, for all-dielectric TMDC microcavities at room temperature, we expect $g_c$ to be much smaller than $g_x$, since photon losses (affecting the cavity photon linewidth) are much weaker than phonon-induced losses (affecting the exciton linewidth) ~\cite{Wurdack2021,Yun2022}. Hence, herein, we set $g_c=0$ and $g_x=g$.
}


The exciton-photon coupling in Eq.~\eqref{eq:Hamiltonian} gives rise to the complex upper ($U$) and lower ($L$) eigenvalue branches:
\begin{equation}
\label{eq:eigenvalues}
    {E}_{U,L} -i {\gamma}_{U,L} = \langle \tilde{E} \rangle \pm \frac{1}{2}\sqrt{(\Delta-i\delta)^2 + 4(V-i\sqrt{g\gamma_x})^2},
\end{equation}
where $\langle \tilde{E} \rangle = \langle E \rangle-i \langle \gamma \rangle$ is the mean complex eigenvalue with $\langle \gamma \rangle=(\gamma_c+g+\gamma_x)/2$, $\Delta = E_c-E_x$ is the bare energy difference and $\delta = \gamma_c + g - \gamma_x$. The eigenvalues and eigenvectors will simultaneously coalesce at the exceptional point when $V=|\delta|/2$ and $\Delta = 2\sqrt{g\gamma_x}$.
Note that when $g=0$, the non-Hermitian term $-iWW^\dagger$ simply describes the decay rates $\gamma_{c,x}/\hbar$ of the bare (uncoupled) cavity photon and exciton, respectively.

The level and linewidth dynamics for different strengths of $V$ and $g$ are shown in Fig.~\ref{fig1}a,b. {\color{black}(The general behaviour in $\Delta$-$g$ parameter space is discussed in Supplementary Information Section S2)}.
We focus on the strong coupling regime at $V>|\delta|/2$ characterised by energy anticrossings and linewidth crossings. In this regime, the eigenstates correspond to the hybrid exciton-polariton quasiparticles \cite{Weissbuch1992,Microcavities,deng2010,Liu2015}. The corresponding shifts of the polariton energies from the bare exciton and photon energies, defined as $\Delta_{UL} = \mathfrak{Re}({E_U}-{E_L} )- \Delta$, is plotted in Fig.~\ref{fig1}c as a function of $\Delta$. In the purely coherent coupling regime ($g=0$), the energies exhibit level repulsion (positive $\Delta_{UL}$) with a maximum value at resonance ($\Delta=0$) where the linewidths cross.
The level repulsion decreases monotonically with $|\Delta|$, reminiscent of the familiar Hermitian limit, $\gamma_{c,x}, g \rightarrow 0$~\cite{Microcavities}.

The behaviour of the energy levels and linewidths is drastically modified when the dissipative coupling $g$ is introduced. The linewidth crossing shifts towards higher $\Delta$ leaving a linewidth repulsion close to resonance, which leads to the so-called resonance trapping of the long-lived state~\cite{ResonanceTrappingExp}. Remarkably, the energy shift $\Delta_{UL}$ becomes negative for a wide range of $\Delta$, indicating level attraction {\color{black} which peaks at a positive $\Delta\approx 2\sqrt{g\gamma_x}$ before monotonically decreasing.}
This strongly $\Delta$-dependent energy shift is responsible for the anomalous dispersion of exciton polaritons presented in this work.

\begin{figure}
\includegraphics[width=\linewidth]{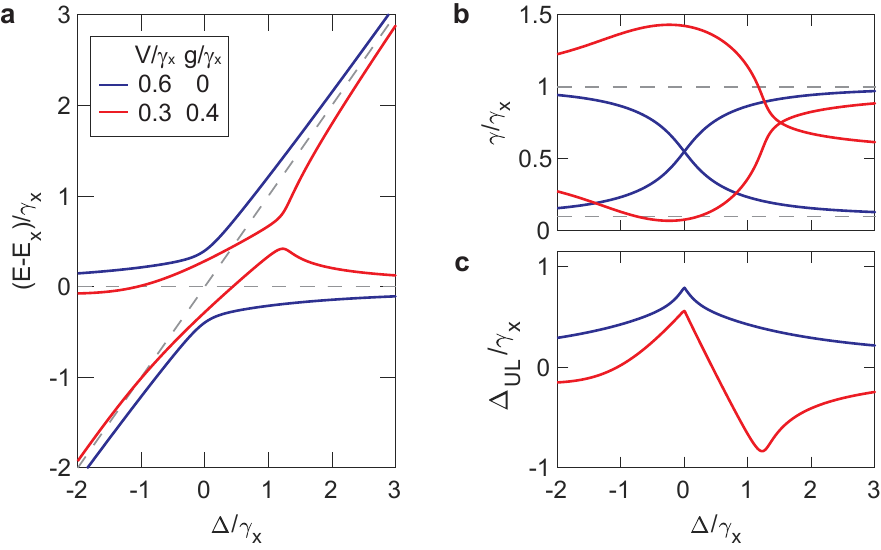}
\caption{\textbf{Coherent and dissipative coupling}. {\bf a}, Energy and {\bf b}, linewidth dynamics for different values of coherent $V$ and dissipative $g$ coupling strengths (in units of $\gamma_x$). 
Dashed lines are the bare energy eigenvalues $E_{c,x}-i\gamma_{c,x}$. {\bf c}, Energy shift $\Delta_{UL}$ (see text) for the parameters used in {\bf a} and {\bf b}. In all cases, we tuned the cavity energy $E_c$ while fixing $E_x$ and $\gamma_{c} = 0.1\gamma_x$.
}
\label{fig1}
\end{figure}

\begin{figure}
\includegraphics[width=\linewidth]{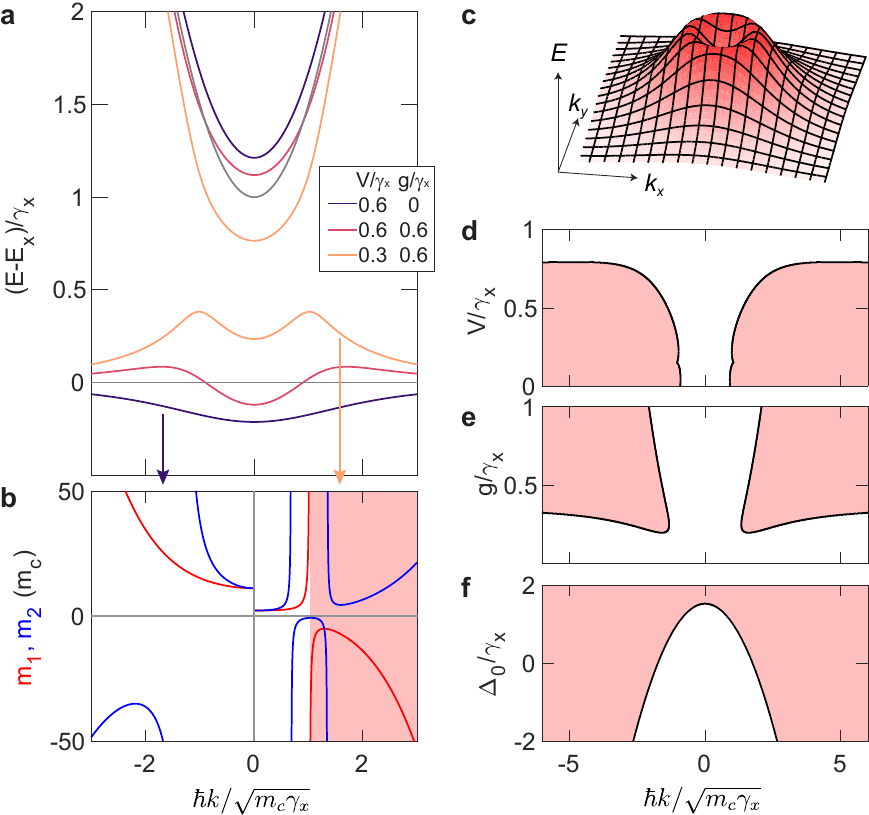}
\caption{\textbf{Anomalous dispersion of exciton polaritons}. {\bf a}, Dispersion for different values of coherent and dissipative coupling strenghs at fixed exciton--photon detuning $\Delta_0 = \gamma_x$. Thin grey lines are the bare exciton $E_x$ and cavity photon $E_c$ dispersions. {\bf b}, $k$-dependent mass parameters $m_1$ (red) and $m_2$ (blue) corresponding to the two dispersions in {\bf a} (see arrows). {\bf c}, Lower polariton dispersion in two-dimensional $k$-space corresponding to {\bf a} with $V=0.3\gamma_x$ and $g=0.6\gamma_x$. {\bf d-f}, Negative-mass ($m_1<0$) regions (shaded) in $k$-space as a function of ({\bf d}) $V$ for $g = 0.6\gamma_x, \Delta_0 = \gamma_x$, ({\bf e}) $g$ for $V = 0.6\gamma_x, \Delta_0 = \gamma_x$, and ({\bf f}) $\Delta_0$ for $V=0.3\gamma_x, g=0.6\gamma_x$.
}
\label{fig2}
\end{figure}

To describe the exciton--photon dispersion, we approximate the cavity photon dispersion as \cite{Microcavities} $E_{c}(k)=E_{0} + \hbar^2k^2/2m_c$, where $E_{0}$ is the cavity resonance energy at normal incidence, $\hbar k$ is the momentum along the plane of the cavity, and $m_c$ is the in-plane effective mass of the cavity photon. The exciton energy $E_x$ is approximately constant within the relevant momentum range probed here. Typical dispersion curves at a positive exciton--photon detuning, $\Delta_0=E_{0}-E_x$, are presented in Fig.~\ref{fig2}a. Without the dissipative coupling ($g=0$), the dispersion features repelling branches corresponding to the well-known upper and lower exciton polaritons~\cite{Microcavities},
where the lower branch is always redshifted from the exciton line.
With increasing dissipative coupling strength $g$, level attraction starts to dominate, firstly starting at higher $k$. When $g$ is large enough compared to $V$, the entire lower polariton branch is blueshifted from the exciton line.

Note that, for this set of parameters, the strongest level attraction occurs at a finite $k$, {\color{black} i.e., where $\Delta(k) \approx 2\sqrt{g\gamma_x}$ and the level attraction peaks (see Fig.~\ref{fig1}c). However, for a very large exciton-photon detuning, i.e. $\Delta_0>2\sqrt{g\gamma_x}$, the strength of level attraction monotonically decreases for all $k$, resulting in a single inversion peak at $k=0$ (see Supplementary Information Section S3).}

To characterise the dispersion, we define the mass parameters~\cite{colas2018} $m_1(k) = \hbar^2 k [\partial_k E(k)]^{-1}$, which determines the group velocity $v_g = \hbar k/m_1$, and $m_2(k) = \hbar^2 [\partial^2_k E(k)]^{-1}$, which determines the acceleration due to an external field. Note that $m_1$ is only negative around the inverted dispersion whereas $m_2$ switches signs at the inflection points. The masses are plotted in Fig.~\ref{fig2}b for the lower branches in Fig.~\ref{fig2}a (indicated by arrows) with and without dissipative coupling. When $g=0$, $m_1$ is positive for all momenta and $m_2$ is only negative at finite $k$, a known feature of exciton polaritons~\cite{colas2018}. This is in stark contrast to the case with $g\neq0$, where both masses become negative near the inversion peak. While $m_2$ switches back to positive sign, $m_1$ remains negative for the plotted range of momenta, shown by the shaded region in Fig.~\ref{fig2}b. Note that the sign of the $m_1$ and $m_2$ for the upper branch is largely unaffected by $g$. 

It is important to point out that the inverted dispersion is isotropic, forming a ring in $k$-space, as shown in Fig.~\ref{fig2}c. This is distinct from the inverted bands in periodic band structures, where the inversion peaks are localised at high-symmetry points only~\cite{Baboux2018}.

We further analyse the $\Delta_0$-$V$-$g$ parameter space as a function of $\hbar k$ to determine under which conditions the inverted dispersion appears. {\color{black} Notably, the model predicts} that the negative-mass regions disappear when coherent coupling significantly dominates over dissipative coupling, as shown by the plots in Fig.~\ref{fig2}d,e.
Hence, either $V$ has to be decreased or $g$ increased to observe the inverted dispersion. The negative-mass regime also persists for a wide range of the exciton--photon detuning $\Delta_0$, as shown in Fig.~\ref{fig2}f. However, the momentum range corresponding to the negative mass increases with the detuning $\Delta_0$ and the dispersion becomes completely inverted at large positive $\Delta_0/\gamma_x$.
In summary, positive exciton-photon detunings $\Delta_0$ and weaker $V$ relative to $g$ favour the negative-mass regime in our system.


\textbf{{\color{black}Experiment: anomalous dispersion.}} To demonstrate the effects of the anomalous dispersion experimentally, we fabricated {\color{black} several planar microcavities} with integrated monolayer WS$_2$ at positive exciton-photon {\color{black} detunings} and with reduced coherent coupling strength $V$. We achieved this {\color{black}by using} substrate engineering and our recently developed technology for integrating monolayer WS$_2$ into polymethyl-methacrylate (PMMA)/SiO$_x$ spaced planar microcavitites \cite{Yun2022}. {\color{black} The relative exciton oscillator strength, and hence $V$, of the monolayer is weakened after the transfer onto the `low-quality' substrate and further material deposition (see Supplementary Information Section S4).}

\begin{figure*}
\centering
\includegraphics[width=\textwidth]{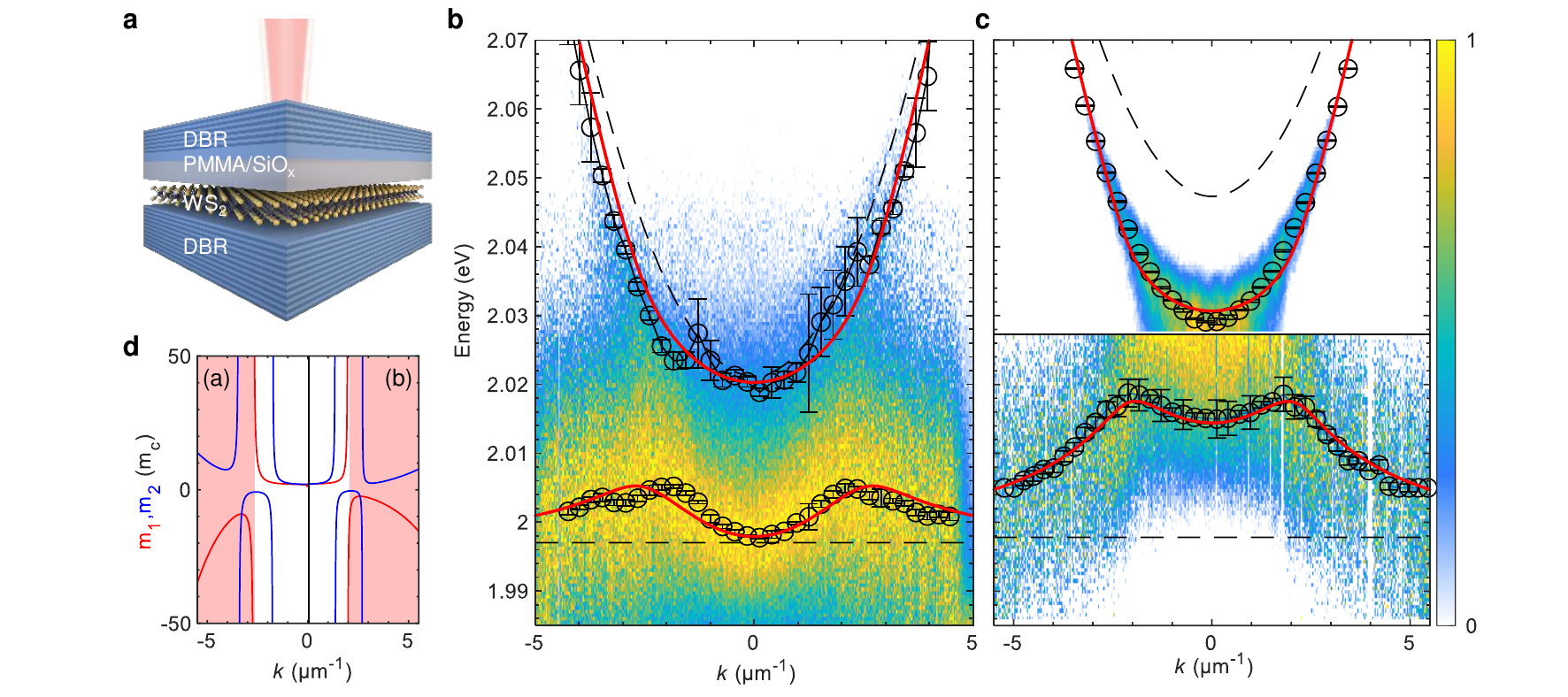}
\caption{\textbf{Experimental realisation of negative mass}. {\bf a}, Schematics of the fabricated microcavity with embedded monolayer WS$_2$. {\color{black} {\bf b,c}, Momentum resolved PL spectra of two samples with different values of the exciton-photon detuning, $\Delta_0$. The maximum intensity at each $k$ value is scaled to unity to visualise the shape of the lower branch. In panel {\bf c}, the normalisation of the spectrum is performed for $E<2.028~\mathrm{eV}$ to suppress the strong emission from the upper branch. The black circles are the fitted peak positions of the two branches. The solid red lines are the fitted dispersion of the upper and lower polaritons and the black dashed lines are the bare microcavity photons and excitons. Error bars in {\bf b,c} represent 95\% confidence interval. {\bf d}, Mass parameters (red) $m_1$ and (blue) $m_2$ of the lower branches in panel (b) and panel (c), with the negative-mass regions shaded in red. }}
\label{fig3}
\end{figure*}

Figure~\ref{fig3}a illustrates the sample design. Here, the WS$_2$ monolayer is mechanically transferred onto the bottom DBR protected with PMMA against further deposition of the thin SiO$_x$ spacer and the top DBR (see Methods). The thin SiO$_x$ spacer allows us to carefully adjust the exciton--photon detuning $\Delta_0$ and the two distributed Bragg reflectors (DBRs) enable a Q-factor of above $10^3$ (see Supplementary Information {\color{black} Section S4}).

The momentum-resolved photoluminescence (PL) spectra of the room temperature polariton emission {\color{black} in two samples with different $\Delta_0$ are shown in Fig.~\ref{fig3}b,c, together with the extracted peak energies of the lower and upper exciton polariton branches.} The observed PL intensities are dictated by the photonic Hopfield coefficients and thermalisation of polaritons \cite{Lundt2016}. Hence, the PL of the upper branch {\color{black} in Fig.~\ref{fig3}c, which is highly photonic at such a large positive $\Delta_0$,} strongly dominates the emission of the structure and decreases at larger energies due to thermalisation. To highlight the emission of the lower polariton branch  {\color{black} in Fig.~\ref{fig3}b,c} we scaled the maximum value of the PL spectrum to unity at each value of $k$. {\color{black} In Fig.~\ref{fig3}c we defined a cut-off energy for normalisation at $E\approx2.028~\mathrm{eV}$ due to the strongly dominating emission from the upper branch.} As seen in Fig.~\ref{fig3}b,c, the lower branch energy decreases towards $E_x$ at large momenta, demonstrating the level attraction shown in Fig.~\ref{fig2}a. {\color{black} We also observe similar behaviour in reflectance measurements (see Supplementary Information Section S6 for more details).}

The complex-valued dispersion branches are extracted by fitting the spectrum at each value of $k$ using a two-peak Voigt function, with the peak energy and linewidth corresponding to the real and imaginary parts of the complex eigenvalues of the system Hamiltonian, respectively {\color{black} (see Supplementary Information Section S5). The $k$-dependence of the extracted peaks were then fitted using the model Eq.~\eqref{eq:eigenvalues}, with the fitting results presented as red solid lines in Fig.~\ref{fig3}b,c. Here, the energies and linewidths of the cavity photons were extracted from the reflectivity measurements of the empty microcavities next to the monolayers (see Supplementary Information Section S4), and fixed for the fits, with the energies shifted by approximately $-8~\mathrm{meV}$ due to the optical thickness of the monolayer. The exciton linewidths and energies are expected to change after the deposition of the top DBR \cite{Yun2022}, thus, these values were chosen as free fitting parameters together with the coherent and dissipative coupling strengths $V$ and $g$. For Fig.~\ref{fig3}b, the fitting yields the values: $\Delta_0 = (24\pm2)~\mathrm{meV}$, $\gamma_x = (31\pm12)~\mathrm{meV}$, $V=(13\pm6)~\mathrm{meV}$ and $g=(18\pm9)~\mathrm{meV}$, and for Fig.~\ref{fig3}c, $\Delta_0 = (51\pm1)~\mathrm{meV}$,  $\gamma_x = (42\pm8)~\mathrm{meV}$, $V=(12\pm5)~\mathrm{meV}$ and $g=(23\pm4)~\mathrm{meV}$.  

The values of the parameters for the two samples are consistent (within errors) and satisfy the condition $2V>|\delta|$, confirming that the samples operate in the strong coupling regime and thus host exciton polaritons. However, due to the suppressed exciton oscillator strength in our samples~\cite{Yun2022} the coupling strength $V$ is smaller than $g$, leading to level attraction and the anomalous dispersion. 
The level attraction at $k=0$ is stronger in the more positively detuned sample. This agrees with the calculations shown in Fig.~\ref{fig1}c, where the magnitude of level attraction increases as $\Delta/(2\sqrt{g\gamma_x})$ approaches unity. {\color{black}Indeed,} $\Delta_0/(2\sqrt{g\gamma_x})$ is $\sim$0.8 ($\sim$0.5) for the more (less) positively detuned sample. 
Since $\Delta_0/(2\sqrt{g\gamma_x})<1$ in these samples, the maximum level attraction occurs at a larger $\Delta$ {\color{black}and $k$}, and therefore, the inversion peaks in these samples are located at a finite momentum, endowing a negative mass $m_1$ to the lower polaritons with momenta $k\gtrsim\pm 2.7~\mu\mathrm{m}^{-1}$ for Fig.~\ref{fig3}b and $k\gtrsim\pm 2~\mu\mathrm{m}^{-1}$ for Fig. \ref{fig3}c.

Note that in samples with the same detuning, $\Delta_0/(2\sqrt{g\gamma_x})$ can be made larger than $1$ by decreasing dissipative coupling $g$ or narrowing the exciton linewidth $\gamma_x$, which results in suppressed level attraction and an inversion peak located $k=0$ (see Fig.~2f).}
{\color{black}
We demonstrated this effect experimentally in a sample made using a microcavity fabrication technique that does not degrade  the oscillator strength of the monolayer exciton and therefore preserves the coherent coupling $V$~\cite{Rupprecht2021,Wurdack2021}, (see Supplementary Information section S7). In this sample, which has the same exciton-photon detuning as that shown in Fig. \ref{fig3}c, we observe much less attraction compared to the samples with weakened $V$ discussed above. This agrees with our model, where the effect of dissipative coupling on the dispersion, i.e., level attraction, depends on the strength of coherent coupling. Nevertheless, the observed level attraction, albeit weak, suggests that dissipative exciton-photon coupling is ubiquitous in monolayer TMDCs but is often screened by the strong coherent exciton-photon coupling.
}




\begin{figure*}[htp]
\centering
\includegraphics[width=\textwidth]{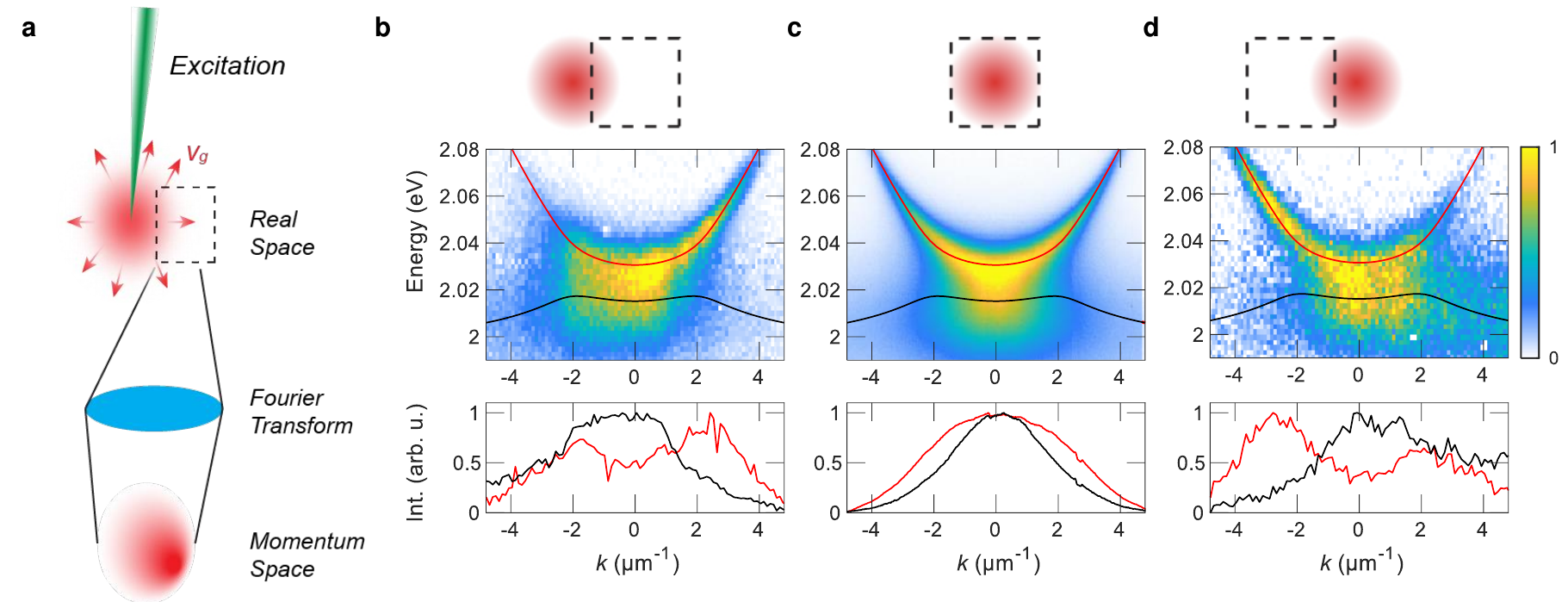}
\caption{\textbf{Negative mass effect on polariton dynamics}. {\bf a}, Schematic of the creation of polaritons in real space (red) by a tight pump laser (green). The polaritons move away from the pump spot with a certain group velocity $\mathbf{v}_g$ (red arrows). A real space filter (dashed box) is used to isolate the detection area in real space before Fourier-transforming to measure the momentum space distribution. {\bf b-d}, Experimental results for different configurations. The upper panel shows the position of the filter with respect to the pump spot (arrows represent the average $\mathbf{v}_g$ along the $x$-direction inside the detection window), the middle panel shows the angle-resolved spectra along $k_x$, and the bottom panel shows the momentum distribution of the (black) lower and the (red) upper branch. The fitted upper and lower polariton branches extracted from Fig. \ref{fig3}b are shown on top of the dispersion measurements for guidance.
}
\label{fig4}
\end{figure*}

\textbf{{\color{black}Experiment: negative-mass dynamics.}} Finally, we show that the anomalous dispersion has a dramatic effect on the dynamics of the lower polaritons {\color{black} using the more positively detuned sample from Fig. \ref{fig3}c}. In contrast to positive mass particles, such as the upper polaritons, negative-mass particles move in the opposite direction to their momentum $\hbar \mathbf{k}$, i.e. their group velocity is opposite to their momentum, as given by the relation $m_1 \mathbf{v}_g = \hbar \mathbf{k}$. To demonstrate this behaviour in the experiment, we excite the sample with a tightly-focused off-resonant laser spot, which results in a spatially localised distribution of polaritons with a wide range of momenta, as shown by the distributions in Fig.~\ref{fig3}b. Polaritons will then move away from the excitation spot in the direction determined by their group velocity $\mathbf{v}_g$ (see  Fig.~\ref{fig4}a). Hence, polaritons displaced  to the positive (negative) $x$ direction with respect to the spot must have an average velocity towards the same direction~\cite{Wurdack2021}. We then measure the momentum distribution of the polaritons displaced from the laser spot by Fourier transforming only a small region in real space (see Fig.~\ref{fig4}b-d) using a spatial filter (see Methods).

Polaritons directly at the excitation spot, measured with a filter centered on the spot (see top of Fig.~\ref{fig4}c), feature a symmetric momentum distribution for both upper and lower branches, as shown by the momentum-resolved PL spectrum (middle panel) in Fig.~\ref{fig4}c. This is clearly seen in the lower panel, where we directly plot the momentum distribution of the two branches by separating the spectrum at $\sim$2.02~eV. {\color{black} To obtain these plots, we integrate the intensities for each panel above and below the cut-off energy, and scale the respective maximum values to unity.} The PL spectrum shows a dramatic change in the distribution for particles displaced to the right of the spot, as shown in Fig.~\ref{fig4}b. The upper polariton emission is skewed to the right, i.e., the average momentum has the same direction as its group velocity, but remarkably, the lower polariton emission is skewed to the left, i.e., its average momentum is opposite to the particle displacement or the group velocity. The additional peak to the left of the  upper polariton distribution likely arises from the tail of the lower polariton distribution, which has a much larger linewidth. This opposite behaviour of momentum and group velocity of the lower polaritons is consistently observed to the left of the laser spot, as demonstrated by Fig.~\ref{fig4}d, where the particles displaced to the left, or with leftward group velocity, have the opposite, rightward average momentum. In all configurations, the  upper polariton branch with the positive mass behaves as expected, i.e. the group velocity and momentum are parallel. Therefore, we have confirmed the negative-mass dynamics of the polaritons using a mixture of particles with distinct signs of effective masses $m_1$. 


\section*{Discussion}

In summary, we have observed dissipative coupling of excitons and photons in monolayer TMDCs (WS$_2$) embedded in planar microcavities at room temperature. This coupling drastically modifies the dispersion of the coupled exciton-photon system, leading to an anomalous (inverted) dispersion for the lower polariton branch. The anomalous dispersion arises when the dissipative coupling between excitons and cavity photons overcomes their coherent coupling, which leads to a negative-mass regime in a large range of momenta at a positive exciton-photon detuning. We have demonstrated the negative-mass effect on the dynamics of polaritons, resulting in the opposing directions of group velocity and momentum. {\color{black} This dynamics should also occur for trion polaritons with anomalous dispersion~\cite{dhara2018, Lyons2022}.}

{\color{black} The dissipative coupling causing the observed effects can arise from the coupling of both excitons and photons to the same decay channel~\cite{ResonanceTrappingExp}, or to a third dissipative mode~\cite{Yu2019}. As discussed above and confirmed by our microscopic theory (see Supplementary Information Section S1), for TMDCs, such as WS$_2$, this coupling likely arises from exciton-phonon interactions~\cite{Li2021Phonon,MoodyLinewidth2015,Christiansen2017}. Recent studies have already shown that phonons play a significant role in exciton--photon interactions~\cite{Li2022polariton}. Furthermore, our hypothesis is supported by the temperature-dependent measurements shown in the Supplementary Information Section S6, where the polariton dispersion exhibits a transition from level attraction to level repulsion by lowering the temperature. This occurs because the exciton-phonon interactions and the relative values of dissipative coupling strength $g$ versus the coherent coupling strength $V$ diminish at lower temperature. Dissipative coupling can also arise via coupling to a `hidden' photon mode~\cite{Yu2019}, but this mode with opposite polarisation is not present in our planar DBR-based microcavities.}

{\color{black} We note that in Ref.~\cite{dhara2018}, a similar level attraction and inverted dispersion was observed for the trion-polariton branch at a large trion-photon detuning. However, a qualitative and heuristic model supporting this finding is not applicable in samples without large doping. We expect doping to be low in our as-exfoliated monolayer integrated into a flip-chip cavity, where an anomalous dispersion is observed at the exciton energy (see Supplementary Information Section S7). In contrast to Ref.~\cite{dhara2018}, our approach explains all observed features by incorporating exciton-phonon coupling and cavity losses, which are always present in these TMDC microcavities and are determined from ab initio calculations. As a result, we would expect that the joint action of cavity losses and exciton-phonon scattering should also contribute to the experimental scenario of Ref.~\cite{dhara2018}, and an effective non-Hermitian Hamiltonian with dissipative coupling could be an appropriate model.}

{\color{black}Various types of dissipative coupling of different origins have been observed in optomechanical~\cite{DissipativeCouplingOpto} and magnon-photon systems~\cite{LevelAttractionMagnon}, and open microwave cavities~\cite{ResonanceTrappingExp, RotterReview}. It also appears in theoretical studies of fragmented exciton-polariton condensates \cite{Rubo2012} and excitons polaritons coupled to an optomechanical resonator~\cite{Kyriienko2014}. Similarly to cavity magnonics and optomechanics, dissipative coupling might be ubiquitous in exciton-photon systems and hence, the anomalous dispersion observed here could potentially also be observed in other materials, e.g. III/V-semiconductors \cite{Weissbuch1992}, perovskites~\cite{Su2020, SuEstrecho2020}, and organic semiconductors~\cite{Plumhof2013}.}



Our work extends the arsenal of dispersion engineering tools for hybrid light-matter particles beyond the application of periodic fields and spin-orbit coupling \cite{Kuehn2010,khamehchi2017negative} by adding the possibility of novel, non-Hermitian disperison engineering.  Here, we only showed the dynamics induced by negative $m_1$. However, the anomalous dispersion can be further employed to demonstrate nontrivial wavepacket dynamics~\cite{colas2018} and negative-mass hydrodynamics~\cite{khamehchi2017negative} due to both negative $m_1$ and $m_2$. For example, the negative-mass polaritons are expected to accelerate in the direction opposite to an applied force. It would be interesting to study bosonic condensation of polaritons in the anomalous dispersion regime, where the energy minimum around $k=0$ (see Fig.~\ref{fig2}c) is not the global one \cite{Ardizzone2022}. The inverted dispersion can also eliminate instabilities of exciton--polariton condensates~\cite{Baboux2018}, which can, for example, enable studies of the Kardar-Parisi-Zhang phase in quantum systems without the complexities of an underlying lattice structure~\cite{KPZPolariton}.


\vspace{0.5cm}
\noindent\textbf{Methods:}
The DBRs are grown by plasma-enhanced chemical vapour deposition (PECVD) and consist of (bottom) 17.5 and (top) 15.5 alternating quarter-wave stack of SiO$_x$ and SiN$_x$, as schematically shown in Fig.~\ref{fig3}a.  Further, the first half of the SiO$_2$ cavity spacer is deposited via RF-sputtering and finished with atomic layer deposition (ALD). The monolayer is then mechanically transferred  at $120\mathrm{^\circ C}$ on top of the oxygen-plasma treated DBR substrate to increase the bonding between the monolayer and the substrate. To protect the monolayer against further material deposition, a $80~\mathrm{nm}$ thick layer of  poly-methyl-methacrylate (PMMA) is spin-coated on top of the structure. Before depositing the top DBR via PECVD to complete the structure, the cavity thickness and hence, the cavity mode, is fine-tuned with an intermediate PECVD grown SiO$_x$ layer. The two DBRs enable a Q-factor of above 10$^3$ (see Supplementary Information Section S4). More details about the fabrication process are reported in Ref.~\cite{Yun2022}. 

The  microcavity is excited with a frequency doubled Nd:YAG laser source at $\lambda=532~\mathrm{nm}$ ($E\approx 2.33~\mathrm{eV}$), which is tightly focused onto the sample surface with a infinity corrected Mitutoyo NIR objective (NA=0.65). The PL is collected with an in-house built optical microscope, which allows for spatial filtering with a square edge-filter prior to momentum-resolved imaging. The momentum-resolved PL-spectra are recorded with an Andor Shamrock 500i spectrograph equipped with an Andor iXon 888 EMCCD camera.

{\color{black}
\vspace{0.5cm}
\noindent\textbf{Data availability:}
The data that support the findings of this study are available from the corresponding authors upon reasonable request.}



\begin{thebibliography}{27}
\bibitem [{ANF()}]{ANFFDBR}%
  \BibitemOpen
  \href@noop {} {}\bibinfo {howpublished} {The fabrication of the
  high-reflectivity substrate was performed at the ACT OptoFab node of the
  Australian National Fabrication Facility - a company established under the
  National Collaborative Research Infrastructure Strategy to provide nano and
  microfabrication facilities for Australia's researchers.}\BibitemShut {Stop}%
\end{thebibliography}
\end{document}


\preprint{topo}




\title{Negative-mass exciton polaritons induced by dissipative light-matter coupling in an atomically thin semiconductor\\
Supplementary Information}



\author{M.~Wurdack}
\email{matthias.wurdack@anu.edu.au}
\affiliation{ARC Centre of Excellence in Future Low-Energy Electronics Technologies and Department of Quantum Science and Technology, Research School of Physics, The Australian National University, Canberra, ACT 2601, Australia}

\author{T.~Yun}%
\affiliation{ARC Centre of Excellence in Future Low-Energy Electronics Technologies and Department of Quantum Science and Technology, Research School of Physics, The Australian National University, Canberra, ACT 2601, Australia}
\affiliation{Department of Materials Science and Engineering, Monash University, Clayton, Victoria, 3800, Australia}
\affiliation{{\color{black}Songshan Lake Materials Laboratory, Dongguan 523808, Guangdong, China}}
\affiliation{{\color{black}Institute of Physics, Chinese Academy of Science, Beijing 100190, China}}

\author{M.~Katzer}
\affiliation{Nichtlineare Optik und Quantenelektronik, Institut f\"ur Theoretische Physik, Technische Universit\"at Berlin,  10623 Berlin, Germany}

\author{A.~G.~Truscott}%
\affiliation{Department of Quantum Science and Technology, Research School of Physics, The Australian National University, Canberra, ACT 2601, Australia}

\author{A.~Knorr}
\affiliation{Nichtlineare Optik und Quantenelektronik, Institut f\"ur Theoretische Physik, Technische Universit\"at Berlin,  10623 Berlin, Germany}

\author{M.~Selig}
\affiliation{Nichtlineare Optik und Quantenelektronik, Institut f\"ur Theoretische Physik, Technische Universit\"at Berlin,  10623 Berlin, Germany}
  
\author{E.~A.~Ostrovskaya}
\email{elena.ostrovskaya@anu.edu.au}
\affiliation{ARC Centre of Excellence in Future Low-Energy Electronics Technologies and Department of Quantum Science and Technology, Research School of Physics, The Australian National University, Canberra, ACT 2601, Australia}

\author{E.~Estrecho}%
\email{eliezer.estrecho@anu.edu.au}
\affiliation{ARC Centre of Excellence in Future Low-Energy Electronics Technologies and Department of Quantum Science and Technology, Research School of Physics, The Australian National University, Canberra, ACT 2601, Australia}
 
\maketitle

\section*{S1: Microscopical Calculations of the polariton dispersion}
In order to self-consistently compute both the polariton dispersion and dephasing, we consider a microscopic Hamiltonian, containing excitons ($X$), phonons ($b$), as well as resonator ($C$) and reservoir photons ($D$), including their individual coupling
\begin{align}
H = H_0 + H_{X-Phon} + H_{X-Phot}  + H_{Phot-Phot},
\end{align}
Here, the free contributions of the different species read
$
H_0 = \sum_{\mathbf{Q}_\parallel
\sigma} E^X_{\mathbf{Q}_\parallel} X^{\dagger\sigma}_{\mathbf{Q}_\parallel} X^\sigma_{\mathbf{Q}_\parallel} + \sum_{\mathbf{K}_\parallel , \alpha} \hbar \omega^\alpha_{\mathbf{K}_\parallel} b_{\mathbf{K}_\parallel}^{\dagger\alpha} b_{\mathbf{K}_\parallel}^{\alpha} + \sum_{\mathbf{\bar Q}_\parallel,\sigma} E_{\mathbf{\bar Q}_\parallel}^C  C_{\mathbf{\bar Q}_\parallel}^{\dagger\sigma} C^\sigma_{\mathbf{\bar Q}_\parallel} + \sum_{\mathbf{Q},\tau} E^D_\mathbf{ Q} D^{\dagger \tau}_\mathbf{Q} D^{ \tau}_\mathbf{Q},
$
where the first term accounts for the dispersion of excitons $E_{\mathbf{Q}_\parallel}^X = E_0^X + \frac{\hbar^2 \mathbf{Q}_\parallel^2}{2 M}$ with the spectral energy $E_0^X$, the mass $M$ \cite{Kormanyos2015} of the exciton and the two dimensional center of mass momentum $\mathbf{Q}_\parallel$. $X^{(\dagger)\sigma}_{\mathbf{Q}_\parallel}$ denote exciton annihilation (creation) operators with the valley $\sigma = K,K' = +,-$, which we assume to commute as bosons in the zero density limit. The second term of $H_0$ denotes the phonon dispersion $\hbar \omega^\alpha_{\mathbf{K}_\parallel}$ obtained from DFT calculations \cite{Li2013,Jin2014}. Furthermore $b_{\mathbf{K}_\parallel}^{(\dagger)\alpha}$ denote phonon annihilation (creation) operators. The third term accounts for the dispersion of the photons in
the cavity $E_{\mathbf{\bar Q}_\parallel}^C = \sqrt{\left(E_0^C\right)^2 + \frac{\hbar^2 c^2\mathbf{\bar Q}_\parallel^2}{(n^C)^2}}$, with the confinement energy $E^C_0 = \frac{\hbar c}{2n^Cd^C}$ adjusted to the experiment (see section S5), the speed of light $c$, the cavity refractive index $n^C$ and the cavity length $d^C$. Photon annihilation (creation) operators in the cavity are denoted by $C_{\mathbf{\bar Q}_\parallel}^{(\dagger)\sigma}$ with the photon polarization $\sigma = \sigma_+ , \sigma_- = +,-$. The last term describes the dispersion of the free photons outside of the cavity $E^D_\mathbf{Q} = \frac{\hbar c |\mathbf{Q}|}{n^D}$.

The exciton-phonon interaction Hamiltonian is given as
$H_{X-Phon} = \sum_{\mathbf{Q}_\parallel,\mathbf{K}_\parallel,\alpha,\sigma} g^\alpha_{\mathbf{K}_\parallel} X^{\dagger\sigma}_{\mathbf{Q}_\parallel + \mathbf{K}_\parallel} X^\sigma_{\mathbf{Q}_\parallel} \left( b^\alpha_{\mathbf{K}_\parallel} +  b_{-\mathbf{K}_\parallel}^{\dagger\alpha} \right)$
with the exciton-phonon matrix element $g^\alpha_{\mathbf{K}_\parallel}$ \cite{Li2013,Jin2014,Selig2022nessy}. 
The exciton-photon coupling Hamiltonian in rotating wave approximation with the excitonic dipole moment $\mathbf{d}$ \cite{Xiao2012,selig2019ultrafast} is given as $H_{X-Phot} = \sum_{\mathbf{Q}_\parallel,\sigma} V_{\mathbf{Q}_\parallel}
    \Big(
    X_{-\mathbf{Q}_\parallel}^{\dagger\sigma} C_{-\mathbf{Q}_\parallel}^\sigma
    +
    X_{\mathbf{Q}_\parallel}^\sigma C_{\mathbf{Q}_\parallel}^{\dagger\sigma}
    \Big)$.

The appearing exciton-photon matrix element reads $V_{\mathbf{Q}_\parallel} = \sqrt{\frac{E^L_{\mathbf{Q}_\parallel}}{\epsilon_0 I^3}} \sin (\frac{\pi z}{d}) \mathbf{e}^\sigma \cdot \mathbf{d}$ (the appearing $\delta_{\mathbf{Q}_\parallel,\mathbf{\bar Q}_\parallel}$ was already accounted for).
In order to describe the out-coupling of cavity photons to the vacuum, we write the phenomenolgical tunneling Hamiltonian 
$H_{Phot-Phot} = \sum_{\mathbf{Q}\sigma\tau} T^{\sigma \tau}_\mathbf{Q} \left( C_{\mathbf{\bar Q}_\parallel}^{\sigma} D_{\mathbf{Q}}^{\dagger \tau}
    +
    C_{-\mathbf{\bar Q}_\parallel}^{\dagger \sigma}  D_{-\mathbf{Q}}^{\tau} \right).$
The Heisenberg equation of motion formalism allows to find a closed set of linear equations, namely the equations of motion (EOM) for the coherent exciton amplitude $\langle X^\sigma_{\mathbf{Q}_\parallel} \rangle$, the coherent cavity photon amplitude $\langle C^\sigma_{\mathbf{\bar Q}_\parallel} \rangle$ and the coherent amplitude of the free photons outside the cavity
$\langle D^\tau_{\mathbf{Q}} \rangle$. To close the system, it is necessary to also compute equations of the respective phonon assisted  amplitudes $\langle X^\sigma_{\mathbf{Q}_\parallel} b^{(\dagger)\alpha}_{\mathbf{K}_\parallel} \rangle$, $\langle C^\sigma_{\mathbf{\bar Q}_\parallel} b^{(\dagger)\alpha}_{\mathbf{K}_\parallel} \rangle$ and $\langle D^\tau_{\mathbf{Q}} b^{(\dagger)\alpha}_{\mathbf{K}_\parallel} \rangle$.
Reflecting the strong coupling between excitons and cavity photons, the whole set of EOM is diagonalised with respect to the coupling $M_{\mathbf{Q}_\parallel}$, which gives $ X^\sigma_{\mathbf{Q}_\parallel} 
    =
    \alpha_{\mathbf{Q}_\parallel}  P^-_{\mathbf{Q}_\parallel\sigma}
    +
    \beta_{\mathbf{Q}_\parallel}  P^+_{\mathbf{Q}_\parallel\sigma}$, and $ C^\sigma_{\mathbf{Q}_\parallel} 
    =
    \alpha_{\mathbf{Q}_\parallel} P^+_{\mathbf{Q}_\parallel\sigma}
    -
    \beta_{\mathbf{Q}_\parallel}  P^-_{\mathbf{Q}_\parallel\sigma}$, with the coherent amplitudes of the (upper and lower) polariton branches $ P^\pm_{\mathbf{Q}_\parallel\sigma}$ and the complex Hopfield coefficients $\alpha_{\mathbf{Q}_\parallel},\beta_{\mathbf{Q}_\parallel} 
    = 
    \frac{1}{\sqrt{2}}
    \sqrt{1
    \pm
    \frac{\tilde\Delta_{\mathbf{Q}_\parallel}}
    {\sqrt{\tilde\Delta_{\mathbf{Q}_\parallel}^2+4 \tilde V_{\mathbf{Q}_\parallel}^2}}}$, with the energy detuning $\tilde\Delta_{\mathbf{Q}_\parallel}
    =
    E^C_{\mathbf{Q}_\parallel}-E^X_{\mathbf{Q}_\parallel}
    -i\big(\gamma^C_{\mathbf{Q}_\parallel}-\gamma^X_{\mathbf{Q}_\parallel}\big)$ and the complex coupling between excitons and cavity photons $\tilde V_{\mathbf{Q}_\parallel} = V_{\mathbf{Q}_\parallel} + ig_{\mathbf{Q}_\parallel}$. Note that both $\tilde\Delta_{\mathbf{Q}_\parallel}$ and $\tilde V_{\mathbf{Q}_\parallel}$ are complex terms, with the real parts stemming directly from the Hamiltonians for dispersion and exciton-photon interaction, respectively. The imaginary parts are however to be found as the solution vector ($\gamma^X_{\mathbf{Q}_\parallel},\gamma^C_{\mathbf{Q}_\parallel},g_{\mathbf{Q}_\parallel}$) of the self consistent EOM, as will be shown in the following. After diagonalisation, the EOM of the phonon assisted quantities can be solved subsequently in a Born-Markov approach, (for $\langle X^\sigma_{\mathbf{Q}_\parallel} b^{(\dagger)\alpha}_{\mathbf{K}_\parallel} \rangle$ even a second diagonalisation is necessary). Eventually, we end up with 
    \begin{align}
	&\Big(i\hbar \partial_t - E^+_{\mathbf{Q}_\parallel} 
    +i\alpha^2_{\mathbf{Q}_\parallel}
	\Gamma^{Pt+}_{\mathbf{Q}_\parallel}
	+i\beta^2_{\mathbf{Q}_\parallel}
	\Gamma^{Pn+}_{\mathbf{Q}_\parallel}
	\Big) 
	\langle P^+_{\mathbf{Q}_\parallel\sigma} \rangle \nonumber\\
	&
	-
	i
	\alpha_{\mathbf{Q}_\parallel}
	\beta_{\mathbf{Q}_\parallel}
	\big(
	\Gamma^{Pt-}_{\mathbf{Q}_\parallel}
	-
	\Gamma^{Pn-}_{\mathbf{Q}_\parallel}
	\big)
	\langle P^-_{\mathbf{Q}_\parallel\sigma} \rangle 
	= 
	\beta_{\mathbf{Q}_\parallel}
    \Omega^\sigma_{\mathbf{Q}_\parallel}\label{eq:P+}\\
	&\Big(i \hbar \partial_t - E_{\mathbf{Q}_\parallel}^- 
	+i\beta^2_{\mathbf{Q}_\parallel}
	\Gamma^{Pt-}_{\mathbf{Q}_\parallel}
	+i\alpha^2_{\mathbf{Q}_\parallel}
	\Gamma^{Pn-}_{\mathbf{Q}_\parallel}
	\Big) 
	\langle P^-_{\mathbf{Q}_\parallel\sigma} \rangle 
	\nonumber\\
	&-
	i
	\alpha_{\mathbf{Q}_\parallel}
	\beta_{\mathbf{Q}_\parallel}
	\big(
	\Gamma^{Pt+}_{\mathbf{Q}_\parallel}
	-
	\Gamma^{Pn+}_{\mathbf{Q}_\parallel}
	\big)
	\langle P^+_{\mathbf{Q}_\parallel\sigma} \rangle 
	= \alpha_{\mathbf{Q}_\parallel}
	\Omega^\sigma_{\mathbf{Q}_\parallel}.\label{eq:P-}
\end{align}
Where $\Omega_{\mathbf{Q}_\parallel}^\sigma$ accounts for the external driving, and the microscopically derived dephasing due to the photonic reservoir outside the cavity reads $\Gamma^{Pt\pm}_{\mathbf{Q}_\parallel}
    =
    \pi
    \sum_{Q_z , \tau\sigma} 
    (T_\mathbf{Q}^{\sigma \tau} )^2
    \delta\big(E_{\mathbf{Q}}^D-E^\pm_{\mathbf{Q}_\parallel}\big)$ and the derived dephasing due to the phonons in the {\color{black} atomically thin transition metal dichalcogenide crystal (TMDC)} can be written as $\Gamma^{Pn\pm}_{\mathbf{Q}_\parallel}
	=
	\pi
	\sum_{\mathbf{K}_\parallel,\alpha}
	|g^\alpha_{|\mathbf{K}_\parallel-\mathbf{Q}_\parallel|}|^2
	\bigg(
	\nonumber\\
	\bar\alpha_{\mathbf{K}_\parallel}^2
	(1 + n_{|\mathbf{K}_\parallel-\mathbf{Q}_\parallel|}^\alpha)
	\delta
	\big(\mathcal{E}^{L}_{\mathbf{K}_\parallel} 
	-
	E^\pm_{\mathbf{Q}_\parallel}
	+ 
    \hbar \omega_{|\mathbf{K}_\parallel-\mathbf{Q}_\parallel|}^\alpha 
	\big)
	\nonumber\\
	+ 
	\bar\alpha_{\mathbf{K}_\parallel}^2
	n_{|\mathbf{K}_\parallel-\mathbf{Q}_\parallel|}^\alpha
	\delta
	\big(\mathcal{E}^{L}_{\mathbf{K}_\parallel} 
	-
	E^\pm_{\mathbf{Q}_\parallel}
	- 
    \hbar \omega_{|\mathbf{K}_\parallel-\mathbf{Q}_\parallel|}^\alpha 
    \big)
    \nonumber\\
	+
	\bar\beta_{\mathbf{K}_\parallel}^2
	(1+n_{|\mathbf{K}_\parallel-\mathbf{Q}_\parallel|}^\alpha)
	\delta
	\big(\mathcal{E}^{U}_{\mathbf{K}_\parallel} 
	-
	E^\pm_{\mathbf{Q}_\parallel}
	+ 
    \hbar \omega_{|\mathbf{K}_\parallel-\mathbf{Q}_\parallel|}^\alpha 
    \big)
    \nonumber\\
    +
	\bar\beta_{\mathbf{K}_\parallel}^2
	n_{|\mathbf{K}_\parallel-\mathbf{Q}_\parallel|}^\alpha
	\delta
	\big(\mathcal{E}^{U}_{\mathbf{K}_\parallel} 
	-
	E^\pm_{\mathbf{Q}_\parallel}
	- 
    \hbar \omega_{|\mathbf{K}_\parallel-\mathbf{Q}_\parallel|}^\alpha 
    \big)
	\bigg)$, with the phonon occupation  $n_{\mathbf{Q}_\parallel}^\alpha = \langle b_{\mathbf{Q}_\parallel}^{\dagger\alpha}b_{\mathbf{Q}_\parallel}^{\alpha}\rangle$ computed from a Bose-Einstein statistics for the lattice temperature. We stress that for both the photonic and the phononic dispersion channels, the Fermi-rules in the delta functions secure that all scattering events obey energy conservation with respect to the self-consistently computed exciton-polariton dispersion $E^\pm_{\mathbf{Q}_\parallel}
    =
    \frac{E_{\mathbf{Q}_\parallel}^X
    +
    E_{\mathbf{Q}_\parallel}^C}{2}
    \pm
    \mathfrak{Re}\left(
    \sqrt{\frac{\tilde\Delta_{\mathbf{Q}_\parallel}^2
    }{4}
    +
    \tilde V_{\mathbf{Q}_\parallel}^2}\right)$. (The imaginary counterpart of this dispersion gives the respective dephasing in the diagonalised basis, $\gamma^{\pm}_{\mathbf{Q}_\parallel}
    =
    \frac{\gamma^X_{\mathbf{Q}_\parallel}+\gamma^C_{\mathbf{Q}_\parallel}}{2}
    \mp
    \mathfrak{Im}\left(
    \sqrt{\frac{\tilde\Delta_{\mathbf{Q}_\parallel}^2
    }{4}
    +
    \tilde V_{\mathbf{Q}_\parallel}^2}\right)$.) Eq.~(\ref{eq:P+},\ref{eq:P-}) can be mapped on a phenomenological model, where $\langle P^+_{\mathbf{Q}_\parallel\sigma} \rangle$ and $\langle P^-_{\mathbf{Q}_\parallel\sigma} \rangle$  must be decoupled after diagonalisation, which gives in total four equations for the computation of the dephasing
    \begin{align}\label{eq:numeqs1}
    \gamma^+_{\mathbf{Q}_\parallel}
    &=
    \alpha_{\mathbf{Q}_\parallel}^2
	\Gamma^{Pt+}_{\mathbf{Q}_\parallel}
	+
	\beta^2_{\mathbf{Q}_\parallel}
	\Gamma^{Pn+}_{\mathbf{Q}_\parallel}\\
	\gamma^-_{\mathbf{Q}_\parallel}
    &=
    \beta_{\mathbf{Q}_\parallel}^2
	\Gamma^{Pt-}_{\mathbf{Q}_\parallel}
	+
	\alpha^2_{\mathbf{Q}_\parallel}
	\Gamma^{Pn-}_{\mathbf{Q}_\parallel}\label{eq:numeqs2}\\
	0
	&=
	\alpha_{\mathbf{Q}_\parallel}
	\beta_{\mathbf{Q}_\parallel}
	\big(
	\Gamma^{Pt-}_{\mathbf{Q}_\parallel}
	-
	\Gamma^{Pn-}_{\mathbf{Q}_\parallel}
	\big)\label{eq:numeqs3}
	\\
	0
	&=
	\alpha_{\mathbf{Q}_\parallel}
	\beta_{\mathbf{Q}_\parallel}
	\big(
	\Gamma^{Pt+}_{\mathbf{Q}_\parallel}
	-
	\Gamma^{Pn+}_{\mathbf{Q}_\parallel}
	\big)\label{eq:numeqs4}
\end{align}
This set of equations completely determines the dephasing, which means that its solution gives the dephasing also in the original basis, i.e. $\gamma^X_{\mathbf{Q}_\parallel}$ as the dephasing of the excitons, $\gamma^C_{\mathbf{Q}_\parallel}$ as the respective dephasing of the cavity photons, and a third dephasing $g_{\mathbf{Q}_\parallel}$, which corresponds to the imaginary part of $\tilde V_{\mathbf{Q}_\parallel}$ and is off-diagonal in the original basis of the Hamiltonian.
Those three parts of the dephasing constitute a vector of numerical solutions of Eqs.~(\ref{eq:numeqs1}-\ref{eq:numeqs4}). Since $\Gamma^{Pt\pm}_{\mathbf{Q}_\parallel}
=
\Gamma^{Pn\pm}_{\mathbf{Q}_\parallel}$ is unlikely, we reduce Eqs.~(\ref{eq:numeqs3},\ref{eq:numeqs4}) to the more likely scenario $\alpha_{\mathbf{Q}_\parallel}
\beta_{\mathbf{Q}_\parallel}=0$, which reduces the set of equations to match the dimension of the solution vector ($\gamma^X_{\mathbf{Q}_\parallel},\gamma^C_{\mathbf{Q}_\parallel},g_{\mathbf{Q}_\parallel}$). As exciton-phonon scattering occurs at a range of momenta and energies substantially exceeding that of the light-cone and the light-matter coupling region, $\mathbf{K}_\parallel \gg \mathbf{Q}_\parallel$, we approximate $\mathbf{Q}_\parallel \approx 0$ and  $|\mathbf{K}_\parallel-\mathbf{Q}_\parallel|\approx\mathbf{K}_\parallel$ \cite{Thranhardt2000}, which significantly reduces the computational overhead. This leaves us with the solution vector $(\gamma^X_0,\gamma^C_0,g_0)$.
%

\begin{figure}
    \centering
    \includegraphics[width=\linewidth]{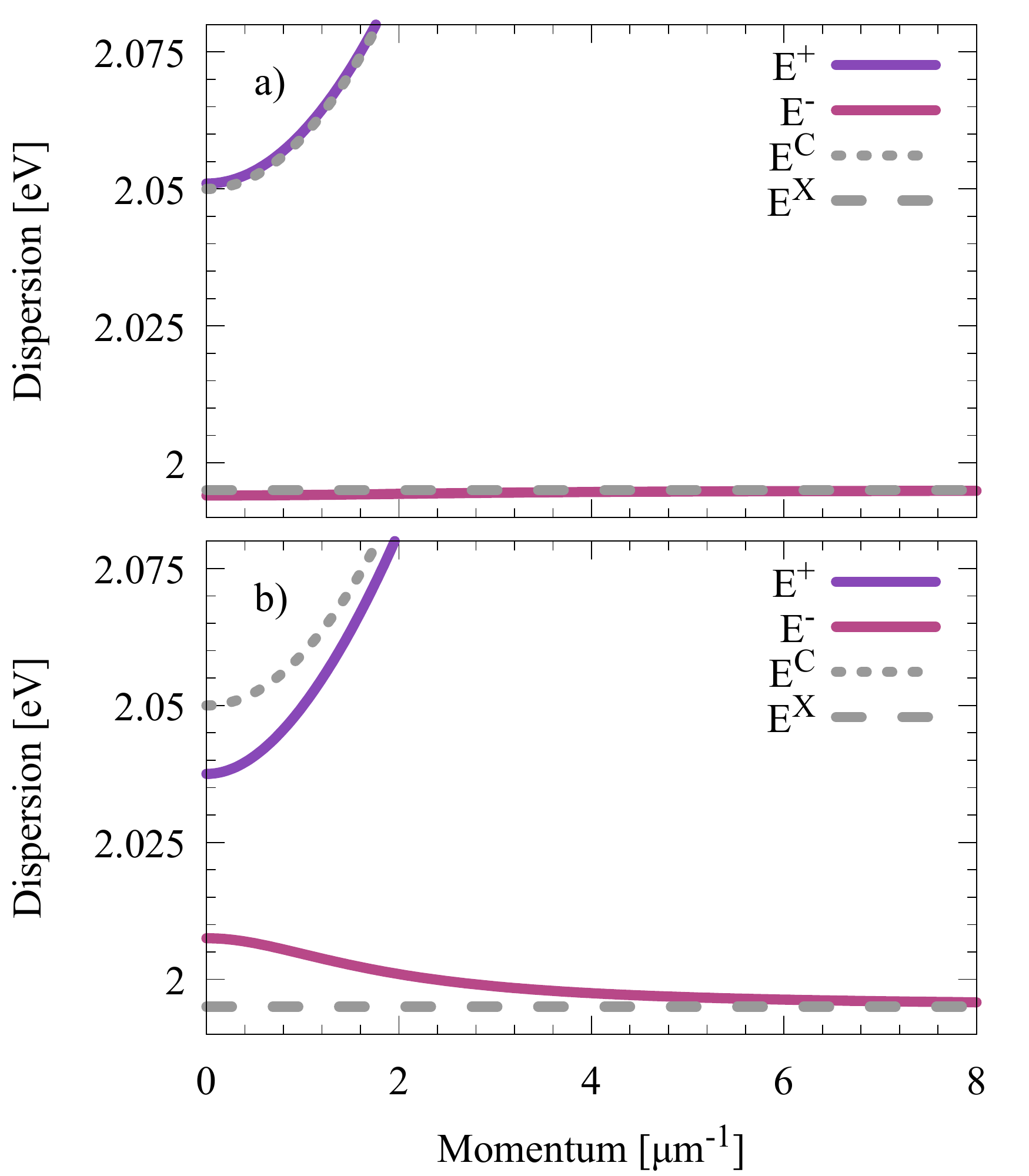}
    \caption{Calculated dispersion of (red) the upper ($E^+$) and lower ($E^-$) polariton branches, and (black dashed) the dispersion of the uncoupled excitons, $E^X$, and of the cavity photons, $E^C$. {\color{black} {\bf a}, Solution without the phonon bath, {\bf b}, solution with the phonon bath.} The level attraction can be traced back to the off-diagonal dephasing $g_{\mathbf{Q}_\parallel}$, which is computed self consistently from the interplay of phononic dephasing in the TMDC and photonic dephasing from the cavity loss. } 
    \label{fig:dispersion}
\end{figure}

Solving this system of equations with {\color{black}$V_0 = \unit[8.3]{meV}$, the exciton-photon coupling strength for $Q_\parallel=0$, $E^C_0=2.05~\mathrm{eV}$, $E^X_0 = 1.995~\mathrm{eV}$, and $T_0 = \unit[0.54]{eV}$ (corresponding to a cavity linewidth of $\Gamma_C = T_0^2/\hbar c_0 \approx \unit[1.47]{meV}$),  we find an off-diagonal dephasing of $g_0\approx \unit[23]{meV}$, which significantly exceeds its real valued counterpart $V_0$, as it is assumed correctly in the phenomenological model in the main text. The microscopically calculated photonic and excitonic dephasings are $\gamma^C_0 \approx \unit[1.8]{meV}$ and $\gamma^X_0 \approx \unit[22]{meV}$, respectively. }

Note, that these solutions are not directly related to the measured linewidths {\color{black} of the uncoupled systems. In particular, $\gamma_0^C$ differs from the cavity linewidth $\Gamma_C$ measured without the presence of the TMDC, and $\gamma_X$ differs from the value one would expect for a TMDC without a cavity surrounding. Both are altered by the strong coupling, and cannot be interpreted without the third dissipation factor arising in this strong coupling regime, which we call off-diagonal dephasing $g_0$. This term relates to the terms $g_c$ and $g_x$ used in the phenomenological model, which represents the dissipative coupling via the photon and phonon baths, respectively. Note however that its value contains contributions from all elements in the $W$ matrix in our phenomenological model.
} 

Fig.~\ref{fig:dispersion} shows the dispersion for the numerically computed values. {\color{black} It is evident that the level attraction is absent when not taking phonons into account (panel a), while with phonons it becomes clearly visible (panel b). This further supports our hypothesis that it is the exciton-phonon coupling that mainly leads to level attraction in our experiment and that the dissipative coupling via the phonon bath is the dominant channel in this material system. Hence, in the phenomenological model, $g_c \ll g_x$, as we used in the main text.}
%

%
{\color{black}  It is important to stress that this microscopic model does not contain any fitting parameters; all applied values rely on ab initio calculations from the literature~\cite{Kormanyos2015,Li2013,Jin2014,Xiao2012}. Since a coupled oscillator Hamiltonian can only produce level attraction (as shown in Fig. S1b) when accounting for off-diagonal dephasing, our microscopic model justifies the parameters $g_c$ and $g_x$ in the phenomenological model, Eqs.~(1,2) in the main text. Therefore, the theory yields a good qualitative agreement between our microscopic calculations, our phenomenological Hamiltonian, and the experiment, with both dissipative coupling and level attraction present.}
{\color{black}
Reproducing the experimentally observed maximum two-level attraction  at $k\neq0$ and formally deriving the phenomenological Hamiltonian from the microscopic model is subject to further studies. }

\section{S2: Energy level dynamics in $\Delta$-$V$ parameter space}
\begin{figure}[h!]
\centering
\includegraphics[width=\linewidth]{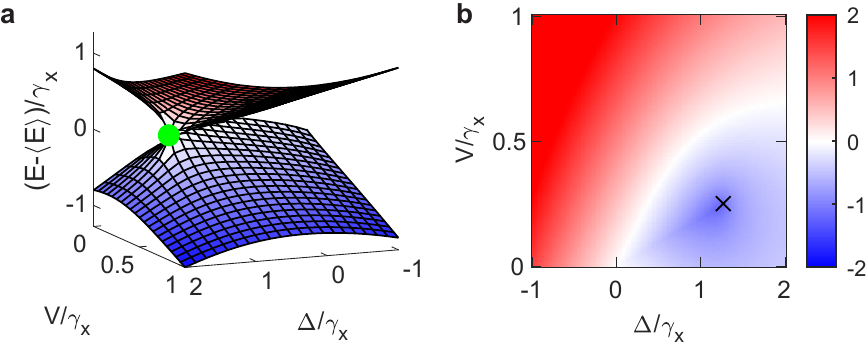}
\caption{\textbf{Exceptional point and energy deviation.} {\bf a}, Mean-subtracted energy surfaces of the eigenvalue branches, and {\bf b}, energy deviation $\Delta_{UP}$ in the $\Delta$-$V$ parameter space. The exceptional point (green dot and `$\times$' mark) is at $V=|\gamma_x-\gamma_c-g|/2$, $\Delta = 2\sqrt{g\gamma_x}$. Other parameters are: $\gamma_c=0.1\gamma_x$ and $g=0.4\gamma_x$.
}
\label{fig:EP}
\end{figure}

The general behaviour of the energy eigenvalues presented in the main text is further described here. {\color{black} Similar to the main text, we apply the approximation $\gamma_c<<\gamma_x$, which holds for TMDs embedded in all-dielectric microcavities at room temperature \cite{Wurdack2021,Yun2022}, and thus, set
$g_x = g$ and 
$g_c=0$ } Figure~\ref{fig:EP}a shows the mean-subtracted energies $E-\langle E \rangle$ in the $\Delta$-$V$ parameter space, highlighting the position of the non-Hermitian degeneracy, the exceptional point. Without dissipative coupling, i.e. when $g=0$, the exceptional point occurs at resonance $\Delta=0$ with the critical coherent coupling strength $V_c=|(\gamma_x - \gamma_c)|/2$. The dissipative coupling term $g$ shifts the exceptional point away from resonance towards $\Delta = 2\sqrt{g\gamma_x}$, while counter-intuitively reducing the critical coherent coupling strength to $V_c=|(\gamma_x - \gamma_c - g)|/2$. This is because $\gamma_c$ is significantly smaller than $\gamma_x$ in these TMD microcavities, so $g$ fills in the difference, resulting in a smaller $V_c$.

Figure~\ref{fig:EP}b shows the energy deviation $\Delta_{UP} = E_{U} - E_{L} - \Delta$ as a function of $\Delta$ and $V$ for $g=0.4\gamma_x$. Regions in red (blue) represent level repulsion (attraction), and the white region represent the transition between them. This clearly shows that level attraction will only become apparent when $V$ is sufficiently small. It is also important to stress that maximum level attraction (negative $\Delta_{UP}$) occurs at the exceptional point, which is marked by `$\times$' in Fig.~\ref{fig:EP}b. This is because the exceptional point (where the coupled energies cross) is shifted away from $\Delta=0$ (where the uncoupled energies cross), resulting in a large energy difference between the coupled and uncoupled energies.

\section{S3: Dispersions at zero and negative exciton--photon detunings}
\begin{figure}[h]
\centering
\includegraphics[width=\linewidth]{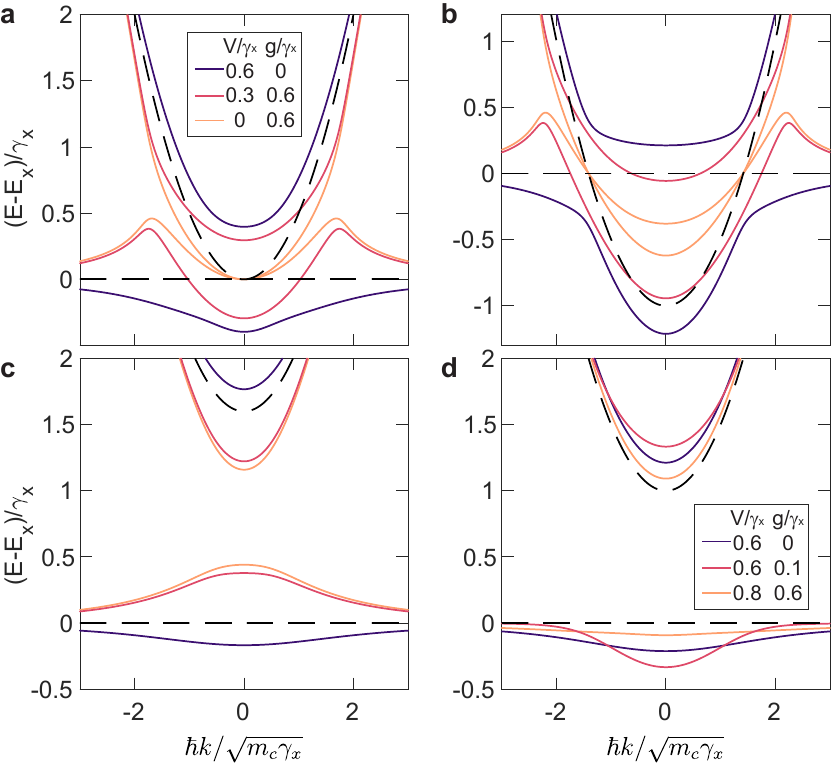}
\caption{\textbf{Dispersions at different exciton--photon detunings.} {\bf a,} Zero ($\Delta_0=0$), {\bf b,} negative ($\Delta_0=-\gamma_x$), and {\bf c,} highly positive ($\Delta_0=2\gamma_x$) detunings. Dashed lines correspond to the exciton and cavity photon dispersions. The values for $V$ and $g$ in {\bf b} and {\bf c} are the same as those in {\bf a}.
}
\label{fig:detuning}
\end{figure}

Figure~\ref{fig:detuning} shows supplementary exciton--photon dispersions at different detunings $\Delta_0$ and coupling strengths $V$ and $g$. The case for resonant ($\Delta_0=0$) detuning is shown in Fig.~\ref{fig:detuning}a, which shows similar features as those in the main text, i.e. an inverted dispersion at finite $k$, when the dissipative coupling strength $g$ is strong. The same features are exhibited by the negative detuning case in Fig.~\ref{fig:detuning}b, except for the crossing in energies when $V=0$, which is expected in the weak coupling regime.

The interesting case of a very positive detuning, with $\Delta_0=2\gamma_x$, is shown in Fig.~\ref{fig:detuning}c, which features an inverted dispersion centred at $k=0$, {\color{black} similar to Fig. \ref{fig:dispersion}b}. Here, both $m_1$ and $m_2$ are negative around $k=0$ with $m_1$ being negative for a wide range of momenta $\hbar k$. Note that in the two cases with $g=0.6\gamma_x$, the lower branch is blueshifted from the exciton line. The parameters here correspond to the upper region of Fig.~2f of the main text.

We also show the dispersion for the set of parameters where there is no negative $m_1$, as presented in Fig.~\ref{fig:detuning}d, corresponding to the upper and lower regions in Fig.~2d and 2e of the main text, respectively. Examples are given for the case when $g$ is present but weak, and when $g$ is strong but $V$ is stronger. In both cases at positive detuning, the lower branch remains below the exciton line.

\begin{figure}[h]
\centering
\includegraphics[width=\linewidth]{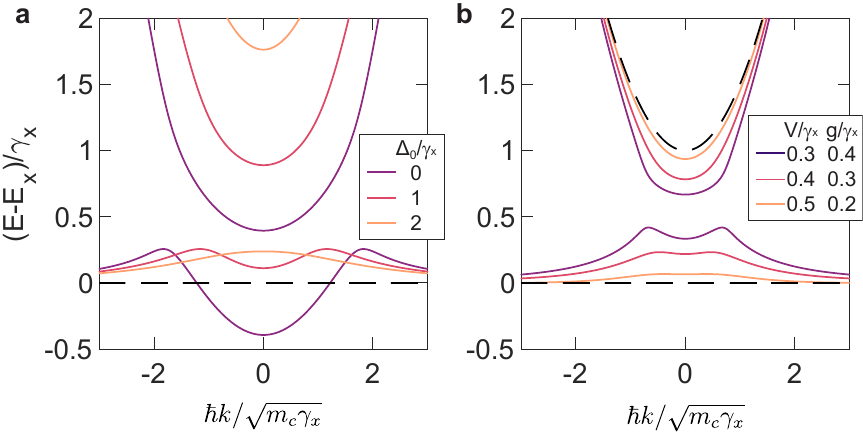}
\caption{\color{black}\textbf{Inverted dispersion for different $\Delta_0$, $V$ and $g$.} {\bf a,} Fixed $V=0.4\gamma_x$ and $g=0.6\gamma_x$. {\bf b,} Fixed $\Delta_0=\gamma_x$. Dashed lines correspond to the exciton and cavity photon dispersions. Bare cavity dispersions not shown in {\bf a} for clarity. The values (from top to bottom of the legend) of $\Delta_0/(2\sqrt{g\gamma_x})$  are ({\bf a}) $[ 0, 0.65, 1.29 ]$ and ({\bf b}) $[0.79, 0.91, 1.12]$.
}
\label{fig:detuning2}
\end{figure}

{\color{black}
Figure~\ref{fig:detuning2}a shows the inverted dispersion as detuning is tuned at fixed $V$ and $g$. One can clearly see that the inverted peak at finite $k$ moves to $k=0$ at high detuning $\Delta_0$, as shown by the map in Fig.~2f of the main text and by Fig.~\ref{fig:detuning}a-c. A similar behaviour occurs if we fix $\Delta_0$ but simultaneously tune the relative strengths of $V$ and $g$.  As shown in Fig.~\ref{fig:detuning2}b, as the $V$ increases and $g$ decreases, the inversion peak moves towards $k=0$. This behaviour corresponds to the difference in dispersion of the two samples shown in the main text (Fig. 3b,c), and what we think occurs when we compare our samples with degraded oscillator strength (see Fig.~3c of the main text) to those with high oscillator strengths (see Fig. S10 and discussion in Section S9).
}

{\color{black}
\section{S4: Optical characterisation of the WS$_2$ excitons and the microcavity photons of the structure presented in the main text.}
}

The structure investigated in the main text was fabricated with the method presented in Ref.~\cite{Yun2022}. In contrast to this work, we deposited the DBR substrate consisting of 17.5 SiN$_x$/SiO$_x$ layers on top of a silicon chip via plasma-enhanced chemical vapour deposition (PECVD) at $300~\mathrm{^\circ C}$, and finalised it with 100$~\mathrm{nm}$ of SiO$_2$, of which 80$~\mathrm{nm}$ was grown via sputtering and $20~\mathrm{nm}$ via atomic layer deposition (ALD). The WS$_2$ monolayer was then mechanically exfoliated \cite{Novoselov2004} from bulk WS$_2$ sourced from HQ graphene \cite{HQgraphene} and placed onto the DBR substrate. As a reference, we placed a WS$_2$ monolayer onto a high-quality SiO$_2$ chip from Nova Materials \cite{Novamaterials}.

After spin-coating the $80~\mathrm{nm}$ thick PMMA cavity spacer and protective layer on both samples, we compare the effects of the substrate on the exciton-photon interactions in these monolayer by measuring the reflectivity spectra with a tungsten halogen white light source.

\begin{figure}
\centering
\includegraphics[width=8.6cm]{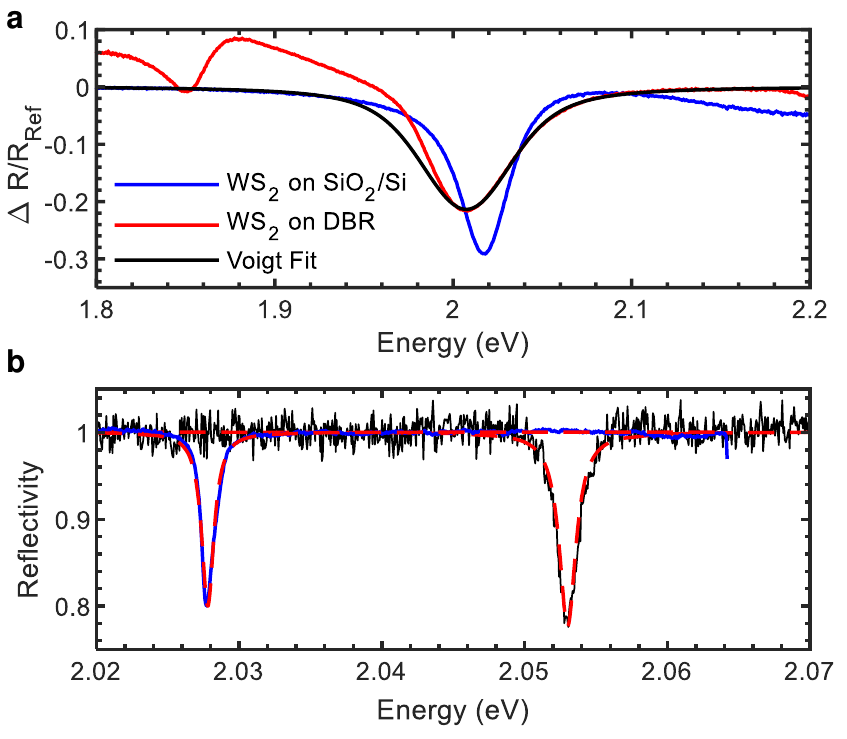}
\caption{\textbf{Characterisation of the microcavity discussed in the main text and the excitonic response of monolayer WS$_2$ on the DBR substrate.} {\bf a}, The derivative of the reflectivity contrast spectrum of a WS$_2$ monolayer capped with PMMA on (blue) a high-quality SiO$_2$ substrate and (red) on the DBR chip. The reflectivity contrast spectrum was fitted with (black) a Voigt profile. {\bf b}, Reflectivity spectra of the microcavities presented in the main text next to the monolayer regions, where (black solid line) is the spectrum of sample 1, (blue solid line) the spectrum of sample 2, and (red dashed line) Lorentzian fits to the respective spectra.}
\label{fig:Fig5}
\end{figure}

Fig.~\ref{fig:Fig5}a presents the reflectivity contrast spectra $\Delta R/R_{Ref}$ of both samples. The spectra were calculated by measuring the reflectivty spectra on the monolayer $R$ and next to the monolayers $R_{Ref}$, with $\Delta R = R-R_{Ref}$. When fitting a Voigt-profile to the high-energy shoulder of the reflectivity spectrum of the monolayer placed on the DBR chip, we find for the exciton energy and linewidth $E_X \approx 2.0073~\mathrm{eV}$ and $\gamma_x= (62.5\pm0.2)~\mathrm{meV}$, respectively{\color{black}, with a homogeneous linewidth of $\gamma_x^H = (34.8\pm0.7)~\mathrm{meV}$ and an inhomogeneous linewidth of $\gamma_x^I = (40.8\pm0.7)~\mathrm{meV}$.}

Here, the product of amplitude and homogeneous linewidth at the exciton energy scales with the exciton oscillator strength quantifying the exciton-photon interactions \cite{Lundt2016}. Clearly, the oscillator strength of the monolayer on the DBR chip is significantly lower compared to that of the reference sample. Therefore, the DBR substrate lowers the exciton oscillator strength of the monolayer compared to that when placed onto a high quality substrate, which is possibly due to environmental charge doping, dielectric disorder and strain \cite{Yu2016,Lippert2017,Raja2019,Khatibi2018,He2013,Sanchez2018}.

Moreover, as we have demonstrated in Ref.~\cite{Yun2022}, the exciton oscillator strength will further decrease after deposition of the top structure, consisting of the SiO$_x$ spacer and the top DBR, via PECVD at $150~\mathrm{^\circ C}$. While a reduction of the exciton oscillator strength is normally avoided for polariton research, it amplifies the effects of dissipative coupling on the polariton dispersion and allows us in this work to observe the negative mass of the lower polariton branch. {\color{black} Additionally, the further deposition of the top structure of the microcavitiy causes a red-shift of the exciton energy \cite{Yun2022}.}

{\color{black} After finalising the two microcavities presented in the main text, we further measured the reflectivity spectra outside of the monolayer areas at zero momentum $k=0$ to characterise the photon mode. Fig.~\ref{fig:Fig5}b shows strong reflectivity dips at $E\approx 2.053~\mathrm{meV}$ and $E\approx 2.028~\mathrm{meV}$ marking the cavity resonances of the two samples $E_c$, which are positively detuned from the bare exciton energy. The linewidths of the cavity modes are $\gamma_c = 1.4~\mathrm{meV}$ for sample 1 and $\gamma_c = 1~\mathrm{meV}$ for sample 2 and therefore, the microcavities have Q-factors of around 1500 and 2000, respectively.}


{\color{black}\section{S5: Peak energies of the angle-resolved PL of the loaded microcavity}

To extract the peak energies of the angle-resolved PL spectra of the monolayers embedded in the microcavity, we fit the spectra at each $k$ with a two-peak Voigt profile. Exemplary fitting results for the highly positively-detuned sample of the main text (see Fig.~3c) are presented in Fig.~\ref{fig:Fig6}. Clearly, the peak energy of the lower branch reaches its maximum value at around $k=2~\mu\mathrm{m}^{-1}$ and decreases at larger $k$ towards the exciton energy (dashed line). This behaviour demonstrates the maximum level attraction at finite $k$, as predicted by our model (see Fig.~1c and Fig2a of the main text), resulting in the anomalous dispersion and negative mass polaritons as presented in the main text (see Fig.~3c).

\begin{figure}
\includegraphics[width=8.6cm]{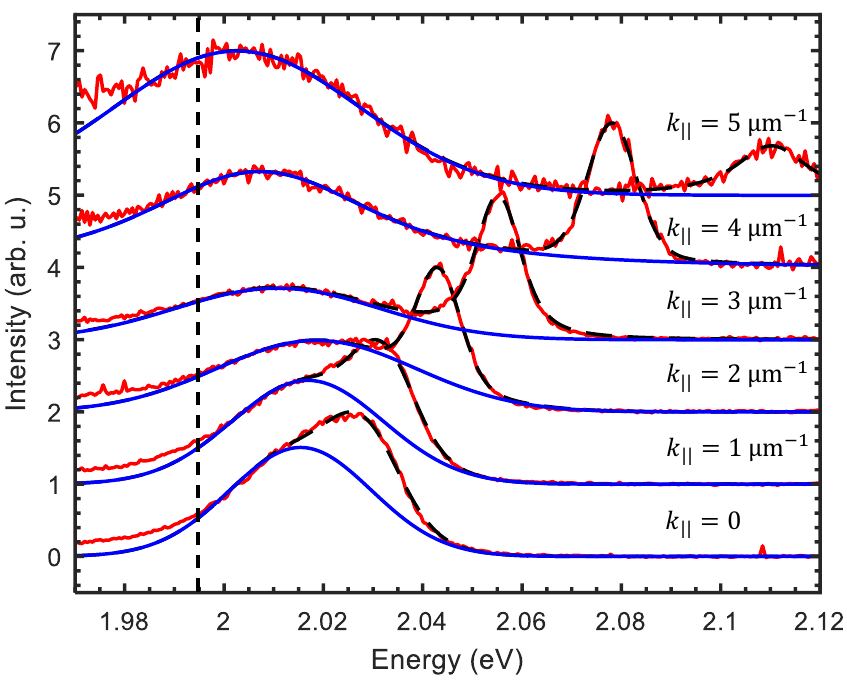}
\caption{\textbf{Waterfall plot of the angle-resolved PL spectrum at increasing momenta $k_{||}$.}  The red lines are the extracted spectra at different $k$, the black dashed lines are the two-peak Voigt fit, and the blue lines are the contribution of the lower branch to the fits. For reference, the fitted exciton energy is marked at $1.995~\mathrm{eV}$. The data correspond to the highly positively detuned sample (see Fig.~3c of the main text).
}
\label{fig:Fig6}
\end{figure}

}

{\color{black}
\section{S6: Reflectance and low temperature measurements}

We fabricated a third sample with the right size and cleaner surrounding areas to enable clear reflectance measurements of the anomalous dispersion. The angle-resolved reflectance contrast and PL spectra of this sample are presented in Fig.~\ref{fig:sample3}a,b. The negative-mass dispersion of the lower branch is visible at high momenta, $k>\pm2.5~\mu\mathrm{m^{-1}}$ and energies below $2.03$~eV. Clearly, the lower branch tends to redshift with increasing $k$ towards the exciton line around $\sim 1.99$~eV and $\sim1.98$~eV for the reflectance and PL, respectively. This is the anomalous dispersion, which is due to the level attraction and hence strong dissipative coupling in this system.

\begin{figure}[h]
\centering
\includegraphics[width=8.6cm]{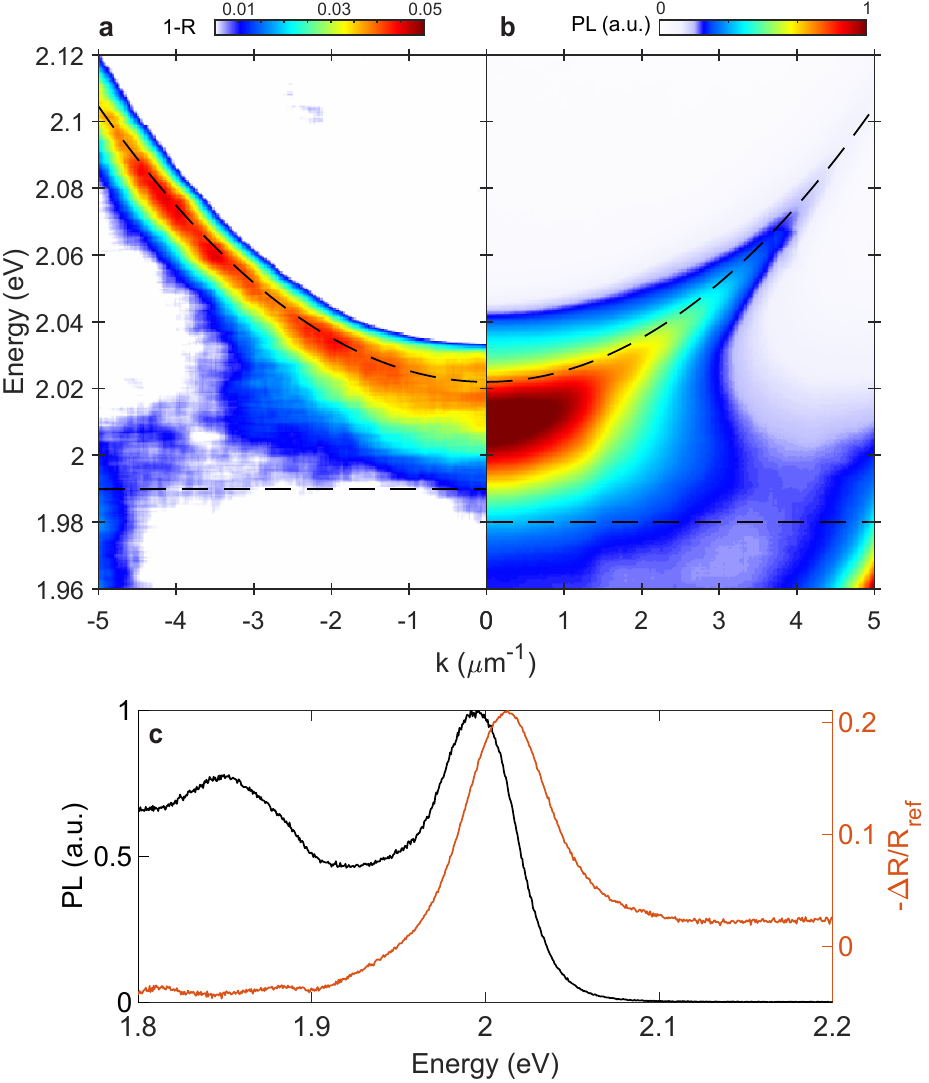}
\caption{\textbf{Characterisation of Sample 3 at room temperature.} Angle-resolved {\bf a}, reflectance contrast and {\bf b}, photoluminescence of a third sample. In \textbf{a}, $R = {\Delta R}/{R_{ref}}$. Dashed lines correspond to the bare exciton and cavity photon dispersion. {\bf c}, PL and negative reflectance contrast $-\Delta R/R_{ref}$ of the monolayer on DBR (before top mirror deposition), showing a Stokes shift of $\approx18$~meV.}
\label{fig:sample3}
\end{figure}

The difference in the bare exciton energies between the reflectance contrast and PL measurements originates from the Stokes shift in this sample. This is shown in Fig.~\ref{fig:sample3}\textbf{c} which were measured on the monolayer on DBR before the top mirror deposition. Note that the further PECVD growth on the structure will redshift the exciton line and change the Stoke's shift as we previously demonstrated~\cite{yun2021influence}. This explains the redshift of the exciton lines when the monolayer is fully embedded in the microcavity.

The anomalous dispersion observed in the reflectance contrast (or absorption) measurements further confirms that the anomalous dispersion observed in the PL measurements correspond to a mode with dissipative coupling and is not a result of excitonic-fraction dependent Stokes shift of the polariton emission, as it was proposed in previous work \cite{Lyons2022}.

To demonstrate tunability on the same sample, we cooled down the third sample to liquid He temperature. At 4~K, the exciton line in our sample blueshifts by around $\sim90$~meV and its linewidth drastically narrows \cite{Jadczak2017}. Furthermore, we expect the exciton-phonon coupling, and hence, the dissipative coupling to be much weaker compared to values at room temperature. Therefore, the level attraction would be diminished as the level repulsion (or coherent coupling) becomes dominant.

\begin{figure}
\centering
\includegraphics[width=8.6cm]{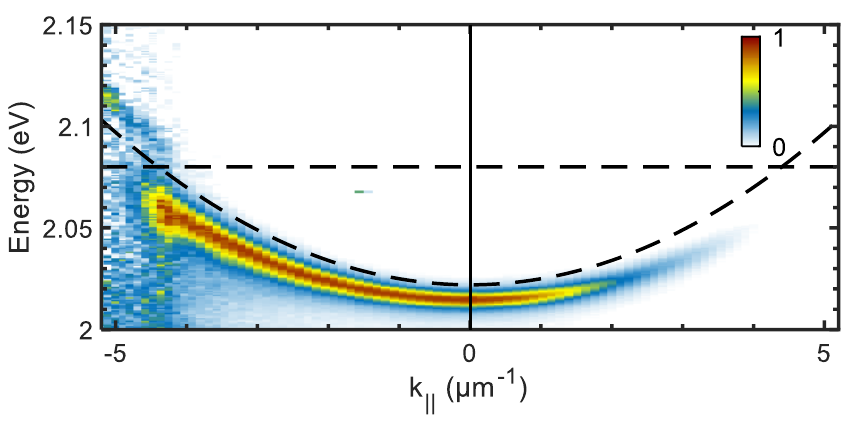}
\caption{\textbf{PL of Sample 3 at low temperature.} Normalised (left) and raw (right) angle-resolved PL spectra of Sample 2 at 4~K with the (black dashed line) uncoupled exciton and photon modes. A clear anti-crossing is observed at the crossing between the exciton and photon modes at   $\sim2.08$~eV.}
\label{fig:sample3_LT}
\end{figure}

This is indeed what we observe in the angle-resolved PL spectra shown Fig.~\ref{fig:sample3_LT}. The exciton-photon detuning at this temperature is very negative due to the strong blueshift of the exciton line $\sim90$~meV. A clear anti-crossing, and thus, level repulsion, is observed around $\sim2.08$~meV which is the main signature of strong coherent coupling between excitons and photons. Hence, we have transitioned from a system dominated by level attraction (dissipative coupling) to that dominated by level repulsion (coherent coupling) on the same sample by tuning the temperature. This is further evidence that the dissipative coupling observed in our samples is indeed a result of exciton-phonon scattering as derived in the microscopic theory (see section S1). A proper study of the effects of temperature on level attraction (or dissipative coupling) requires a sample, in which we can dynamically adjust the cavity energy while reducing the temperature to maintain a constant exciton-photon detuning. This can be done, for example, with an open cavity device, e.g., \cite{Lyons2022}, and is subject to further studies.

\section{S7: Comparison with a flip-chip cavity with a high-quality DBR substrate}
As highlighted in the main text, the quenching of the coherent coupling term and the subsequent domination of the dissipative coupling term arises mainly from the microcavity fabrication process and the quality of the DBR. To support this postulate, we fabricated another sample, we call Sample 4, using a flip-chip approach \cite{Rupprecht2021} and a high-quality DBR substrate~\cite{ANFFDBR}, see Fig.~\ref{fig:sample4}a. We have used this approach before to create high-quality exciton polaritons \cite{Wurdack2022, Wurdack2021} so we expect that coherent coupling is much stronger in this sample.

\begin{figure}
\centering
\includegraphics[width=8.6cm]{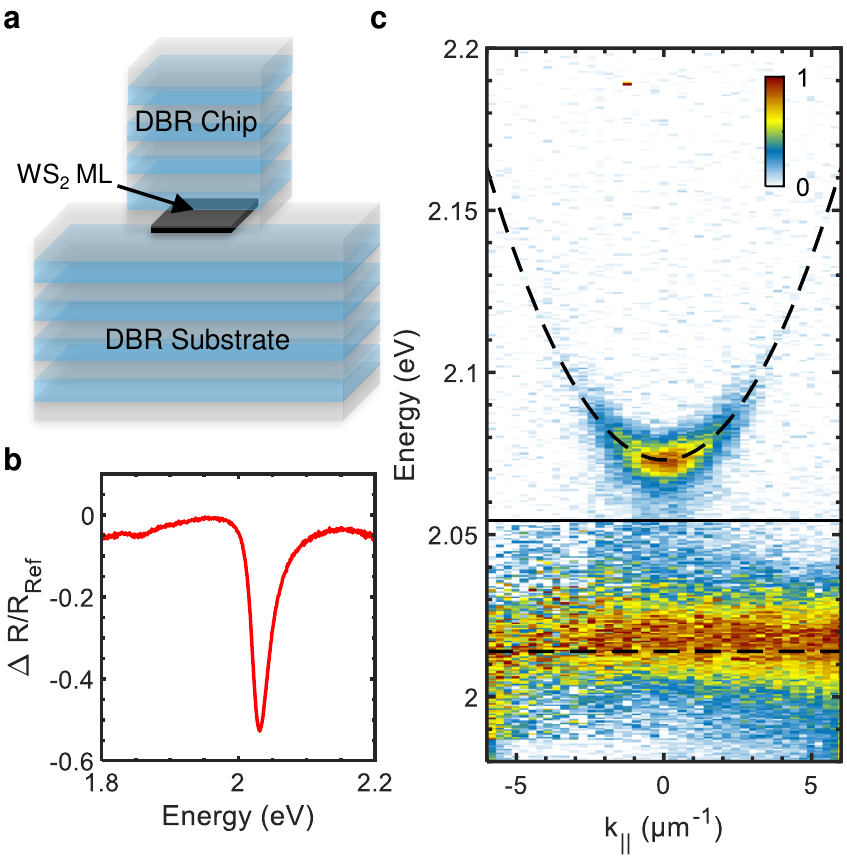}
\caption{\textbf{Flip-chip microcavity with a positive exciton-photon detuning.} {\bf a}, Schematics of the microcavity design with an integrated monolayer (ML) WS$_2$ \cite{Rupprecht2021}. {\bf b}, Normalized reflectivity spectrum of a monolayer WS$_2$ placed onto the high-quality DBR substrate \cite{ANFFDBR}. {\bf c}, Angle-resolved PL spectrum of the structure with the (black dashed lines) uncoupled exciton and photon modes.}
\label{fig:sample4}
\end{figure}

The key difference in this sample is the preservation of the large exciton oscillator strength of the monolayer. In comparison to the monolayer WS$_2$ on PECVD-grown DBR (see ~\ref{fig:Fig5}a), the monolayer WS$_2$ on top of the high-quality DBR maintains a large reflectance dip (see Fig.~\ref{fig:sample4}b), which is proportional to the oscillator strength or the coherent coupling strength. More importantly, this large value is unaffected by the top DBR chip since it is not deposited onto the monolayer \cite{Rupprecht2021}. This is unlike the PMMA-PECVD approach mainly used in this work, which unavoidably encapsulates the monolayer and exposes it to the plasma involved in the PECVD growth.

The resulting PL spectra of Sample 4 is presented in Fig.~\ref{fig:sample4}c showing a well separated upper and lower branches due to the large positive exciton-photon detuning. Interestingly, the lower branch shows a small negative curvature which becomes clear when compared with the flat exciton line. While the detuning in this sample matches that of Sample 1 (with the PL shown Fig. 3c) the level attraction is much smaller in this sample. This result is experimental evidence that increasing the coherent coupling strength leads to reduced level attraction and thus, effects of dissipative coupling on the polariton dispersion as predicted by our model (see also Fig.~\ref{fig:detuning2}). 

Note that similar signatures of weak level attraction of excitons (or trions) and photons resulting in negative-mass dispersions were observed on MoSe$_2$ by other groups \cite{Dhara2018,Eunice2022} but their interpretation is not based on dissipative coupling. As correctly predicted by our model (see Fig.~1c and Fig.~2f of the main text), this attraction is revealed at large positive detunings $\Delta_0>2\sqrt{g\gamma_x}$.
These multiple observations of the weak level attraction (including from Sample 4) from several groups, e.g., \cite{Dhara2018,Eunice2022}, along with the theoretical calculations presented in this work, strongly suggest that dissipative exciton-photon coupling is ubiquitous in monolayer TMDCs but is often screened by the strong coherent exciton-photon coupling. 
}

\bibliography{bibliography}